\documentclass[manuscript]{aastex}

\shorttitle{A New Sample of Distant Compact Groups From DPOSS}
\shortauthors{Iovino et al.}

\begin{document}

\title{A New Sample of Distant Compact Groups From DPOSS}

\author{A. Iovino}
\vskip -0.2 truecm
\affil{Osservatorio Astronomico di Brera, Via Brera 28, I-20121 Milano, Italy}
\email{iovino@brera.mi.astro.it}

\author{R.\ R. de Carvalho\altaffilmark{2}}
\vskip -0.2 truecm
\altaffiltext{2}{Currently at Osservatorio Astronomico di Brera, Via Brera 
28, 20121 - Milano, Italy}
\affil{Observat\'orio Nacional, Rua General Jos\'e Cristino, 77, 20921-400, S\~ao Crist\'ov\~ao,
Rio de Janeiro, Brazil}
\email{reinaldo@on.br}

\author{R.\ R. Gal} 
\vskip -0.2 truecm
\affil{Johns Hopkins University, Dept. of Physics and
Astronomy, 3701 San Martin Dr., Baltimore, MD 21218}
\email{rrg@skysrv.pha.jhu.edu}

\author{S.\ C. Odewahn} \affil{Arizona State University, Dept. of
Physics \& Astronomy, Tempe, AZ 85287} \email{Stephen.Odewahn@asu.edu}

\author{P.\ A.\ A. Lopes}
\vskip -0.2 truecm
\affil{Observat\'orio Nacional, Rua General Jos\'e Cristino, 77, 20921-400, S\~ao Crist\'ov\~ao,
Rio de Janeiro, Brazil}
\email{paal@on.br}

\author{A. Mahabal, S.\ G. Djorgovski}
\vskip -0.2 truecm
\affil{Palomar Observatory, Caltech, MS105-24, Pasadena, CA 91125}
\email{aam@astro.caltech.edu, george@astro.caltech.edu} 

\begin{abstract}
We have identified eighty-four small, high density groups of galaxies
out to $ z\sim 0.2$ in a region of $\sim 2000$ square degrees around
the north galactic pole using DPOSS (the Digitized Second Palomar
Observatory Sky Survey).  The groups have at least four galaxies
satisfying more stringent criteria than those used by Hickson in his
pioneering work in 1982: the adopted limiting surface brightness for
each group is brighter (24 mag arcsec$^{-2}$ instead of 26 mag
arcsec$^{-2}$), and the spread in magnitude among the member galaxies
is narrower (two magnitudes instead of three). We also adopt a
slightly modified version of the isolation criterion used by Hickson,
in order to avoid rejecting groups with projected nearby faint background
galaxies. A 10\% contamination rate due to projection effects is
expected for this sample based on extensive simulations.

\end{abstract}

\keywords{galaxies: clusters of --- galaxies: surveys --- methods:
statistical}

\section{Introduction}

A large fraction of all galaxies in the universe is found in groups
(Nolthenious \& White 1987).  Groups probe intermediate scales between
isolated galaxies and rich clusters, and are therefore important from
a cosmological standpoint, providing constraints on the density
parameter, $\Omega_0$, and on the spectrum of primordial density
fluctuations (e.g. West, Oemler \& Dekel 1989). The study of groups is
also important for investigating the relevant processes in the course
of galaxy formation and evolution.

Despite their importance and frequency of occurrence, galaxy groups
are much less studied than clusters of galaxies.  Identification of
structures with small numbers of galaxies is a difficult task because
of contamination by the field, motivating the search for high density
contrast systems. This special class of groups, called compact groups,
was first observed by Stephan (1877) and Seyfert (1948).  Later,
extensive searches were conducted using the first Palomar Observatory
Sky Survey (POSS-I, covering the sky north of $\delta$ =
-27$^{\circ}$), allowing a more systematic study of these small galaxy
systems (Vorontsov-Velyaminov 1959, 1977; Arp
1966). Vorontsov-Velyaminov (1959), by visually inspecting POSS-I
plates, produced a list of 355 peculiar and interacting galaxy
systems. Taking advantage of the light-gathering power of the 200-inch
telescope on Palomar, Arp (1966) re-examined these peculiar systems
and found other unusual types, producing an atlas of peculiar galaxies
with 338 objects. These two works have shown the importance of
compiling a more complete list of interacting systems, which could
ultimately be used for studying environmental effects on galactic
properties.  Not as popular, but equally important, is the work of
Shakhbasian (1973), who also visually inspected 200 POSS-I plates,
covering 18\% of the northern sky, and cataloged 376 small galaxy
groups. Unfortunately, the search criteria used in these works relied
on subjective values of critical selection parameters, and used visual
inspection to establish object magnitudes and isolation from nearby
structures. Motivated by the need for a homogeneous and
well-quantified sample of compact groups, several authors produced
objectively defined catalogs which are suitable for statistical
analysis. Rose (1977) was the first to establish a quantitative method
to search for compact groups, covering 7.5\% of the sky and finding
205 systems, including triplets, quartets, and quintets. Although
Rose's work pioneered the systematic study of small galaxy systems,
the best studied sample is that of Hickson (1982), which covers 67\%
of the northern sky and lists exactly 100 groups, with a median
redshift of 0.03.  This is the only sample for which redshift
information is available for every group. More recently, new samples
were defined based on digitized catalogs, avoiding most of the
subjectivity of the previous compilations. Examples include the
catalogs of Prandoni, Iovino \& MacGillivray (1994), and Iovino
(2002), which lists 121 compact groups in the southern sky over an
area of $\sim$5000 square degrees.  Other studies have been done
utilizing redshifts for group identification. This alternative, which
obvously alleviates most of the problems with interlopers that purely
photometric searches encounter, is still limited by the availability
of deep, large area redshift surveys, which are only recently becoming
available (e.g.  Barton et al. 1996, Barton, de Carvalho \& Geller
1998; Allam \& Tucker 2000, Focardi \& Kelm 2002).

The level of AGN and starburst activity among compact group members
was recently measured systematically, and an important relation
established between morphology, density and activity (Coziol et
al. 1998a,b).  A subsample of 17 compact groups in the southern sky
shows an excess of AGNs in the central regions, with the surrounding
areas dominated by starburst activity. This relation has been
confirmed by Coziol, Iovino, and de Carvalho (2000) using a larger
sample, and may represent an important diagnostic measure of the
degree of dynamical evolution in a group.  The selection of a
statistically significant sample of groups, not only in the nearby
Universe, but also at higher redshifts, is of paramount importance to
further test and extend such relations.

Iovino (2002) examined the main pitfalls of the Hickson sample and
presented a new automated algorithm to search for groups. This new
algorithm is especially suited to reach fainter magnitudes and can be
properly adapted to extend the search for groups to higher redshifts.
We have made use of this algorithm to define a sample of small,
extremely dense groups of galaxies from the new digitized DPOSS survey
catalog.  This catalog has star/galaxy separation reliability of 90\%
at $r = 19.5^m$ (Gunn-Thuan system). Moreover, it provides color
information $(g-r)$ which can be valuable in understanding the
properties of the sample selected. Within an area of $\sim 2000$
square degrees around the north galactic pole, we selected a sample of
84 small, extremely compact groups, singled out by their high surface
density of member galaxies, high density contrast with respect to the
neighboring galaxy distribution, and narrow spread in magnitude among
the member galaxies. These criteria provide an efficient way to
minimize contamination by background/foreground galaxies, a main
concern when going to fainter magnitudes.

In a hierarchical cosmological scenario, larger mass objects form from
the non-linear interactions between smaller mass units (Press \&
Schechter 1974).  Group progenitors would be formed during the early
clustering of galaxies (or galaxy progenitors) and so one possibility
to identify them and study their dynamical evolution is to extend
current samples to higher redshifts.  As the DPOSS catalog is
released, we intend to build an objective sample of small, dense
groups of galaxies covering the entire northern hemisphere out to $z
\sim 0.2$.  This sample will help investigate how interactions and
mergers, likely to occur in such extreme environments, influence
galaxy evolution.

This paper is organized as follows: \S2 presents a brief description
of the data used for the group search, methods employed in the group
definition, and the group catalog. In \S3 we discuss the statistical
significance of this new catalog, and in \S4 we summarize the main
points presented in the paper, as well as the prospects for future
studies of small galactic systems in the slightly larger redshift
domain we probe here.
                            
\section{Data and Algorithm Used}

Our search for Compact Groups is based on the DPOSS catalogs, which
contain information for stars and galaxies in 3 colors (photographic
$JFN$ bands, calibrated using CCDs to Thuan-Gunn $gri$, and corrected
for Galactic extinction), down to a limiting magnitude of
$r\sim21^m$.  The star-galaxy classification is accurate to $>90\%$ at
$r\le 19.5^m$ (Odewahn et al. 2002). For more details on how the
catalogs were defined and their photometric quality we refer the
reader to Gal et al. (2002). The details of the star-galaxy separation
procedures are found in Odewahn et al. (2002).

We conducted our search using DPOSS catalogs covering a modest area
of $\sim 2000$ square degrees around the north galactic pole. Our
input catalog contains only galaxies in the magnitude range $16.0^m
\leq r \leq 20.0^m$, and we restrict our search to groups whose
brightest galaxy has $16.0^m \leq r \leq 17.0^m$.

The main goal of our algorithm is to keep the contamination by
background/foreground galaxies down to reasonable levels, a difficult
task in the magnitude range considered.  The increase in surface
density of galaxies seriously hampers the search, and the algorithm
should maximize the probability of selecting groups which are real
physical configurations.

As the algorithm will be run on a catalog of galaxies, we use a
very simple tool to judge the amount of contamination in our group
catalog by chance projection effects.  We can apply the algorithm to
select candidate groups from random galaxy catalogs, generated by
shuffling, plate by plate, the coordinates of galaxies (ra,dec) in our
original catalog, while preserving the magnitude and color information
for each galaxy (see Prandoni, Iovino, \& MacGillivray 1994).  This
way, any inherent correlation between color and magnitude of the
catalogue galaxies is preserved.  The plate by plate approach was
chosen in order to preserve differences, sometimes on the order of
30\%, in galaxy surface density among different plates.

The shuffled and the original galaxy catalogs will have the same
distribution of galaxy apparent magnitudes and colors, while the
spatial correlation is lost through the shuffling.  The shuffled
database is thus the parent sample in luminosity and colors for the
compact groups and represents a good mixture of field and cluster
galaxies.  The candidate compact groups that are obtained applying our
algorithm to such a shuffled catalog can be viewed as chance
projection groups, as they are detected because of projection effects
in a catalog where the spatial distribution of galaxies is
randomized. We repeated the generation of shuffled galaxy catalogues
twenty times, and used the mean value of the number of chance
projection groups detected, $\it N_{random}$. In the extreme case of
the presence of high density contrast structures in the galaxy
distribution within single plates, we could be over-estimating the
number of expected groups, as the structure in question will be only
partially washed out by the shuffling of the coordinates.  On the
other hand, we note that shuffling was done on the full plate
catalogue down to magnitude 20$^{m}$, and at such faint magnitudes
large, high density contrast structures are rare.

The ratio of the number of chance projection groups selected
from the shuffled galaxy catalog to the number of 
compact group candidates selected from the
original database, is a good estimate of the
contamination rate (defined as $\it N_{random} / \it N_{real}$) of our
final catalog. 

According to the criteria originally devised by Hickson, an
aggregation of galaxies is defined as a compact group if it contains four
or more galaxies in a compact configuration and is isolated from the
surrounding field (H82).  The criteria are expressed by the following
formulae:

\begin{itemize} 

\item{{\bf richness}: $n \; \ge \; 4$ , where $n$
       is the number of galaxies within 3 magnitudes of the
       brightest group member ($m_{brightest}$).}

\item{{\bf isolation}: $R_{isol} \; \ge \; 3R_{gr}$,
      where $R_{gr}$ is the radius of the smallest circle
      containing the centers of the group members and $R_{isol}$ is
      the distance from the center of the circle to the nearest
      non--member galaxy within 3 magnitudes of the brightest group
      member.}

\item{{\bf compactness}: $\mu_{gr} \; < \; \mu_{limit}$, where
      $\mu_{gr}$ is the mean surface brightness (magnitudes
      arcsec$^{-2}$) within the circle of radius $R_{gr}$.} 

\end{itemize} 
 
All magnitudes refer to the red E band of POSS-I, with which Hickson performed
his visual search. In this photometric band $\mu_{limit} = 26.0$
magnitudes arcsec$^{-2}$.

At fainter magnitudes, the surface density of the sample of galaxies
considered for the search increases. Background galaxies can, by
chance, be superimposed on the isolation ring of a foreground group,
causing it to be rejected by the isolation criterion.  The reverse can
also happen: a faint galaxy can be projected onto an existing triplet,
causing it to be accepted by the richness criterion.

At brighter magnitudes, the surface density of the faintest galaxies
entering the sample is so low that these two problems do not greatly
affect the search.  Furthermore, in a visual search there are other
(implicit) criteria used to select the sample, which can help
alleviate these problems. As shown in Prandoni Iovino, \& MacGillivray (1994), 
the sample selected visually at bright magnitudes by Hickson satisfies
tighter constraints than those stated explicitly.  However, in an
automated search targeting fainter magnitudes, these two problems can
become a major source of both incompleteness and contamination. We
investigate, with the help of randomized catalogs, how modifying the
H82 criteria can minimize the contamination by chance projections.

We begin by exploring how changing $\mu_{limit}$ (the surface
brightness limit imposed by the compactness criterion) to brighter
values, while keeping the other parameters as in H82, affects
contamination rates. Following the approach described above, we
estimate the contamination rate as a function of $\mu_{limit}$ by
running our search algorithm with different values of $\mu_{limit}$,
both on multiple random galaxy catalogs and on the original catalog. We
estimate, for different values of $\mu_{limit}$, the number $\it
N_{random}$ of groups expected to be found purely due to chance
projection effects, and compare it to the number $\it N_{real}$ of
groups found in the original catalog.

The ratio $\it N_{random} / \it N_{real}$ is $\sim 27$\% when
restricting the search to groups with $\mu_{limit}$ $\leq 24$.  Using
fainter surface brightness limits, the contamination rate increases
dramatically, reaching $\sim 50$\% for $\mu_{limit}$ $\leq 25$, and
$\sim 80$\% for $\mu_{limit}$ $\leq 26$.  We therefore set the surface
brightness limit to $\mu_{limit}$ = 24 mag arcsec$^{-2}$, in order to
reduce the contamination rate of our sample to reasonable values. The
strong dependence of contamination rates on the value of $\mu_{limit}$
is due to the high surface density of galaxies in the fainter
magnitude ranges explored in our search.  Only the highest density
contrast configurations are truly improbable at these limits, and we
restrict our search to these configurations only.

We proceed, maintaining $\mu_{limit}$ = 24 mag arcsec$^{-2}$, to
investigate how narrowing the magnitude range among group member
galaxies affects the contamination rate.  We run our algorithm on real
and randomized catalogs and, similar to the approach taken for
$\mu_{limit}$, test how the ratio $\it N_{random} / \it N_{real}$
varies as a function of $\Delta mag_{comp}$, the magnitude range
allowed between the brightest and the faintest member galaxies.
Adopting $\Delta mag_{comp} \leq 2.5^m $, the contamination rate drops
to $\sim 16$\%, as compared to $\sim 27$\% with $\Delta mag_{comp}
\leq 3^m $, as in H82.  A further narrowing of the range to $\Delta
mag_{comp} \leq 2^m$ reduces the contamination rate to $\sim 10$\%.
We thus adopt this final value for the richness criterion.

Tightening the interval of magnitude considered for richness, and
choosing a bright surface brightness limit, reduces the contamination
rates at fainter magnitudes. In other words, selecting extreme
configurations, both in terms of local density and magnitude
concordance among members, is a good strategy to avoid being
overwhelmed by projection effects at fainter magnitudes. We note that
tightening the compactness criterion is merely an acknowledgement of
what is already known to be true also for the H82 sample - the
galaxies in the groups present in his catalogue are actually within a
tighter magnitude range than that stated explicitely (his catalogue is
incomplete when the magnitude range among member galaxies is $>2$; see
Prandoni, Iovino, \& MacGillivray (1994).

For the isolation criterion, we follow a strategy similar to that
described in Iovino (2002). Adopting an absolute isolation criterion in
our search would result in the rejection of a large fraction of group
candidates, which are apparently not isolated because of projection
effects. We adopt a {\it flexible} isolation criterion, such that no
galaxy of magnitude brighter than $m_{faintest}$+0.5$^m$ is within the
isolation radius, where $m_{faintest}$ is the magnitude of the
faintest group member.  Relaxing the isolation criterion in this way
alleviates the incompleteness problem due to the high surface density
of fainter galaxies, while still rejecting cases of
subcondensations within larger structures. The usefulness of the
strict isolation criterion adopted by H82 has already been questioned
in the literature.  A detailed redshift study of 17 groups from H82
shows that most of them are part of larger structures, and only a few
cases can be defined as truly isolated (de Carvalho et al. 1997,
Ribeiro et al. 1998).  Only a dedicated redshift survey of the
surrounding fainter field population can truly assess the actual
degree of isolation of the groups selected.

Thus, our final criteria for the compact group selection algorithm are:
\begin{itemize} 

\item{{\bf richness}: $n \; \ge \; 4$ in the magnitude
       interval $\Delta mag_{comp} = m_{faintest}-m_{brightest}$, with
       the constraint $\Delta mag_{comp} \leq 2^m $.} Group members are 
       those defined by this criterion. 

\item{{\bf isolation}: $R_{isol} \; \ge \; 3R_{gr}$, where
      $R_{isol}$ is the distance from the center of the circle to the
      nearest non--member galaxy within 0.5 magnitudes of the
      faintest group member.}

\item{{\bf compactness}: $\mu_{gr} \; < \; \mu_{limit}$, where
      $\mu_{gr}$ is the mean surface brightness (magnitudes
      arcsec$^{-2}$) within the circle of radius $R_{gr}$, and
      $\mu_{limit}=24.0$ in $r$ band.}

\end{itemize} 

Each candidate group is therefore characterized by 5 quantities: (1)
the magnitude of its brightest galaxy $ m_{brightest}$ ; (2) the
number of its members, $ n_{memb}$; (3) the magnitude interval between
its brightest and faintest member $\Delta mag_{comp}$; (4) the
magnitude interval for which the group is isolated, $\Delta
mag_{isol}$; and (5), its surface brightness, $\mu$. It should be
noted that with our choice for richness and isolation we avoid using
the magnitude bin $19.5^m \leq r \leq 20.0^m$, where star--galaxy
separation accuracy drops from 90\% to 85\%. For each group, only
galaxies in the range $ m_{brightest} < m < m_{faintest} + 0.5 $ are
considered for selection purposes.

The choice of extreme values for the selection criteria, while
considerably reducing the effects of contamination, does not result in a 
paucity of local analogues of such systems.  Out of the one-hundred 
groups selected in H82 sample, thirty have $\mu$ brighter than our limit,
a high percentage which confirms the incompleteness of H82 at fainter
surface brightness limits.  Interestingly, only one of these is 
due to projection effects, while six are triplets and the rest higher
multiplicity systems. Twelve H82 groups that further satisfy the
criterion $\Delta mag_{comp} \leq 2^m $ are all quartets except for one
triplet. The future availability of the Sloan database, with good
star-galaxy separation even in the bright magnitude regime, will allow us
to select a reliable, local comparison sample, following the same
algorithm adopted here.

\subsection{Contamination rate tests}
Considering the importance of establishing a reliable contamination
rate measure, we performed two specific tests in order to gain some
insight into this issue. Therefore, we considered: (a) a purely random
distribution; and (b) a Rayleigh-Levy (RL) distribution. 

An uniform random distribution was generated by choosing a random pair
of (ra,dec), within the limits of a plate, and associating a pair of
magnitude and color (selected from the original galaxy catalog) to
it. This was done plate by plate to preserve the correct galaxy
surface density. The results obtained for the purely random catalog are
consistent with those using the shuffling scheme. In the truly
random case we find 7.1$\pm$3.9 expected groups, while in the shuffled
case we find 8.2$\pm$2.8 expected groups. In both cases the procedure
was repeated fifteen times, and the rms quoted results from the
scatter among different simulations. It should be noted, however, that
this procedure only estimates a lower limit to the true contamination
rate in our sample, as it considers only truly random projections of
galaxies in a uniform Universe, missing other instances of ``fake''
tight groupings on the sky. These include projections of galaxies
within larger structures (e.g., within loose groups, see Mamon 1986,
Mamon 1987 and Diaferio et al. 1994; or along walls, or filaments, see
Hernquist, Katz, \& Weinberg 1995), and real pairs or triplets with
additional unassociated galaxies projected onto them.

To address some of these issues, we generated and  RL distribution 
(see Peebles 1993) which reproduces the
observed two-point correlation function of galaxies, and our group search
was performed on this new galaxy database. The number of groups
obtained in this case is only slightly higher than in the case of the
purely random distribution, totalling nine in the simulation we
performed. This result confirms that neglecting clustering in the real
data does not result in a large error in our "chance projection
groups" estimate. 

A more realistic way of measuring the contamination rate would
be to consider a large cosmological simulation which implicitly
assumes that we ``know'' what the correct model Universe is, but such a 
simulation is outside the scope of this paper. Our
contamination rate estimate should be taken as an indicative
value.  Our main goal is not to establish an absolute contamination
rate but rather to explore how the relative contamination rate changes
as a function of the different parameters adopted for our search, and
how properties of these ``fake'' groups differ from those of the real
candidate groups.

We also experimented with searching for triplets applying the same
criteria stated above (with the obvious change to $n \; \ge \; 3$ of
the richness parameter). The contamination rate for systems with three
members only turned out to be exceedingly high (of the order of 60\%
in the magnitude range of our search), suggesting that, unfortunately,
 the search for these interesting systems is highly inefficient when no 
redshift information is available.

\subsection{The compact group catalog} \label{catalog} 

Running the algorithm described in the previous section on our galaxy
database, we obtain a list of 84 compact group candidates (hereafter
called PCG groups). We verified each candidate by eye, to ensure that
no obvious plate defect was classified as a group member. As
incompleteness of the galaxy database used could also be a concern, we
also check each group candidate for the possible presence of a galaxy
(missing in the database) in the isolation ring.

Table 1 lists the main parameters for each PCG group. Column headings
are as follows:
\begin{enumerate}
\item{Group name.}
\item{Right Ascension (J2000), the coordinates of the
center of the smallest circle containing all member galaxies.}
\item{Declination (J2000).}
\item{Group radius ($R_{gr}$) in arcminutes, the radius of the
smallest circle containing the member galaxies.}
\item{$r$ magnitude of the brightest group galaxy, $m_{brightest}$.}
\item{Mean surface brightness $\mu_{gr}$ of the group ($r$ mags
arcsec$^{-2}$), measured as the total magnitude of the member galaxies
averaged over the area defining the group. The compactness constraint
imposes $\mu_{gr} < 24.0 $.}
\item{Magnitude interval between the brightest and the faintest group member 
($\Delta mag_{comp}$). The richness constraint imposes $\Delta mag_{comp} \leq
2.0$.}
\item{Magnitude interval between the brightest group
member and the brightest interloper in the isolation ring ($\Delta
mag_{isol}$).  The isolation constraint imposes $\Delta mag_{isol} \geq
\Delta mag_{comp} + 0.5^m$.  This quantity is set to 99 when no interloper 
is found within the isolation ring.}
\item{Number of group members, $n$. The richness
constraint imposes $n \geq 4$ within an upper limit of 2 magnitudes
from the brightest group member ($m_{brightest}$).}
\end{enumerate}

Table 2 provides information on the member galaxies of each PCG group.
Column headings are as follows:
\begin{enumerate}
\item{Galaxy name, as shown in Figure 2. All member galaxies are labeled in 
alphabetical order starting from the brightest and going to fainter members.}
\item{Galaxy right ascension (J2000).}
\item{Galaxy declination (J2000).}
\item{$r$ magnitude.}
\item{$g-r$ color.}
\item{Position angle.}
\item{Ellipticity.}
\end{enumerate}

\section{Statistical Significance of the New Sample}

Extensive random simulations have been performed, as described in the
previous section, applying the same selection criteria to different
random realizations of the DPOSS database.  These simulations have
shown that with the criteria chosen we expect 10\% contamination due
to projection effects. In Figure 1, we compare various properties of
our sample with those of ``fake'' groups, selected by the same
algorithm, but using the randomized DPOSS database.  All histograms
have been normalized to one, to emphasize the difference in
shape. Panel (a) shows hints of a depletion of fainter member galaxies
in real groups with respect to the ``fake'' ones, an effect that could
be related to the dense environment (e.g. cannibalism, secular
evolution). Panel (b) shows the surface brightness distribution in our
sample compared to that in ``fake'' groups. The real groups have
higher surface brightnesses with respect to the ``fake'' groups. In
our automated search we are not biased to be incomplete as a function
of magnitude interval between member galaxies, nor of surface
brightness, and therefore these differences probably point to real
evolutionary effects. Finally, panel (c) shows the spread of $g-r$
colors within our sample and within ``fake'' groups, demonstrating a
color concordance within each group. This can be interpreted as a
further check of the physical reality of the groups in our sample: we
expect that real group member galaxies will share the same redshift
and same environmental influences, resulting in similar colors among
member galaxies.  

In Figure 2, we present the DPOSS $F$ (red) images of our entire
sample. In each finding chart the circle drawn has radius $R_{group}$,
while the horizontal bar to the lower left corresponds to a length of
one arcminute.  All member galaxies are labeled in alphabetical order
starting from brightest to faintest.  Some of the images shown are
reminiscent of the chain galaxies seen at higher redshifts (Cowie, Hu
\& Songaila 1995), and could be useful local analogues which can be
studied in greater detail. Our catalog can be used to assess the
density of such objects in the relatively nearby universe.

Figure 3 shows two groups from H82 which are confirmed to be at
redshifts 0.14 and 0.07, respectively. The two images were extracted
from the same plate material used to produce Figure 2 (and the same
that, once cataloged, was the database for our search) and the
resemblance to the groups of our sample is impressive. Interestingly,
 these two Hickson groups would both have
been selected by our new, more stringent, criteria. As noted earlier, the 
few groups from
H82 which are found in the magnitude range of our search satisfy much
tighter constraints than those stated explicitly by Hickson.

\section{Summary}

We have used the DPOSS catalog of galaxies in an area of $\sim 2000$
square degrees around the north galactic pole to search for small
groups out to $z = 0.2$. Our sample consists of 84 groups with an
expected contamination rate of 10\%, based on the simulations
described earlier.

We will enlarge our sample to eventually cover the full northern sky
available to DPOSS (over 10000 square degrees), where we expect a
total of $\sim 400$ candidates.  We can estimate the space density of
the groups in our catalog by assuming that the median redshift of our
sample is $z \sim 0.1$, based on the magnitude distribution of the
brightest group galaxies. This yields a space density of $\sim1.9\times
10^{-5} h^{3}~Mpc^{-3}$ groups (for $\Omega_{M}=1$ and
$\Omega_{\lambda}=0$). This is in good agreement with the results of
Mendes de Oliveira \& Hickson (1991), who quote a space density of
$3.9\times 10^{-5} h^{3}~Mpc^{-3}$ for the H82 sample, especially since we
use more stringent selection criteria in our search.  Obviously, this
order of magnitude estimate needs to be assessed more firmly once
redshift information becomes available.  Spectroscopic follow-up of
our catalog will allow confirmation of these systems as soon-to-merge
 structures, as well as enable the study of galaxy merger rates,
complementing the picture emerging from studies of close galaxy pairs
by including physical systems of higher multiplicity.  The objective
criteria by which our sample is defined will ease the comparison with
N-body simulations and samples at different redshift ranges.

As shown in Figures 2 and 3, our group candidates closely resemble the
compact groups defined by Hickson in his pioneering paper. Our
candidates represent more distant counterparts of the highest density
systems in H82, and their detailed study will allow us to study the
evolutionary history of such structures. Signs of galaxy evolution
have already been reported for clusters and groups above redshift
$\sim 0.1$ (the Butcher-Oemler effect, see Carlberg et
al. 2001). Detecting evolutionary effects for our high density groups
would provide further observational constraints for theories and
models of the processes that drive galaxy evolution at low and
intermediate redshifts.

\acknowledgments 
We thank the Norris Foundation and other private donors for their
generous support of the DPOSS project.  We also thank the Palomar
Telescope Allocation Commettee and Directors for generous time
allocations for the DPOSS calibration effort. We would like to thank
Jack Sulentic for several useful discussions throughout this project,
and the referee, who, through several important comments, helped to
improve the presentation of this paper considerably.  This work was
made possible in part through the NPACI sponsored Digital Sky project
and a generous equipment grant from SUN Microsystems. Access to the
POSS-II image data stored on the HPSS, located at the California
Institute of Technology, was provided by the Center for Advanced
Computing Research.

\clearpage

\begin{figure}
\plotone{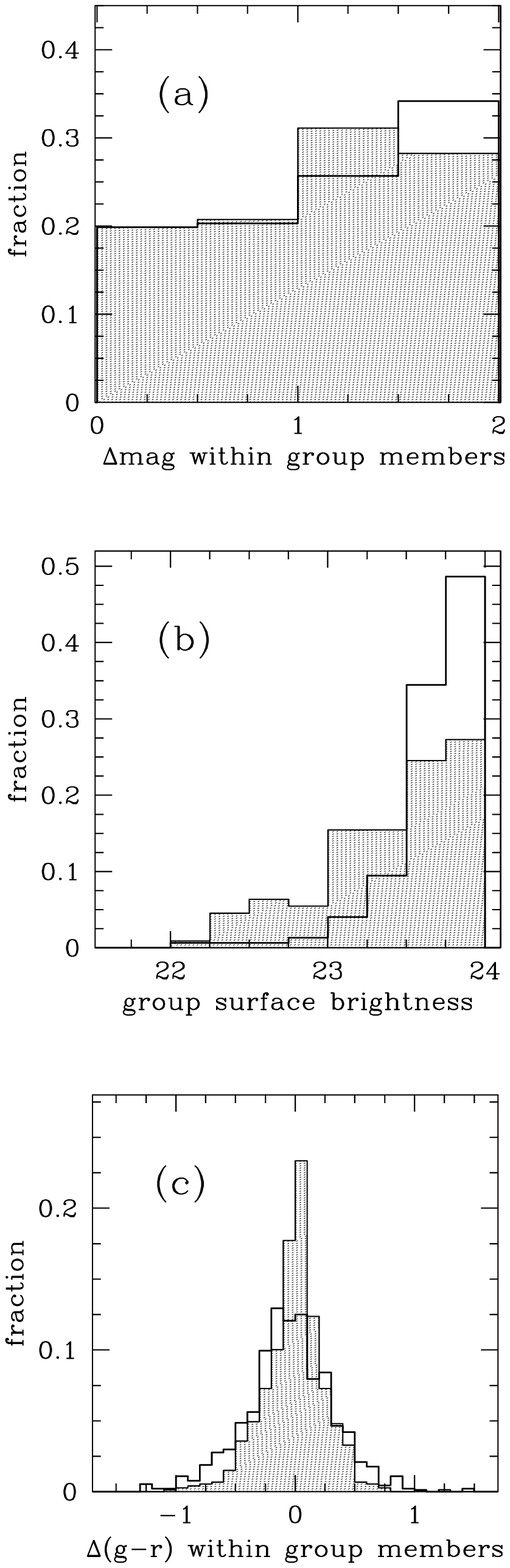} 
\caption{Various properties of our sample groups (shaded
histograms) are compared with those of ``fake'' groups, selected by
the same algorithm in the randomized DPOSS database.  All
distributions have been normalized to one. See text for details.  }
\label{Figure 1}
\end{figure}
\clearpage

\begin{figure}
\plotone{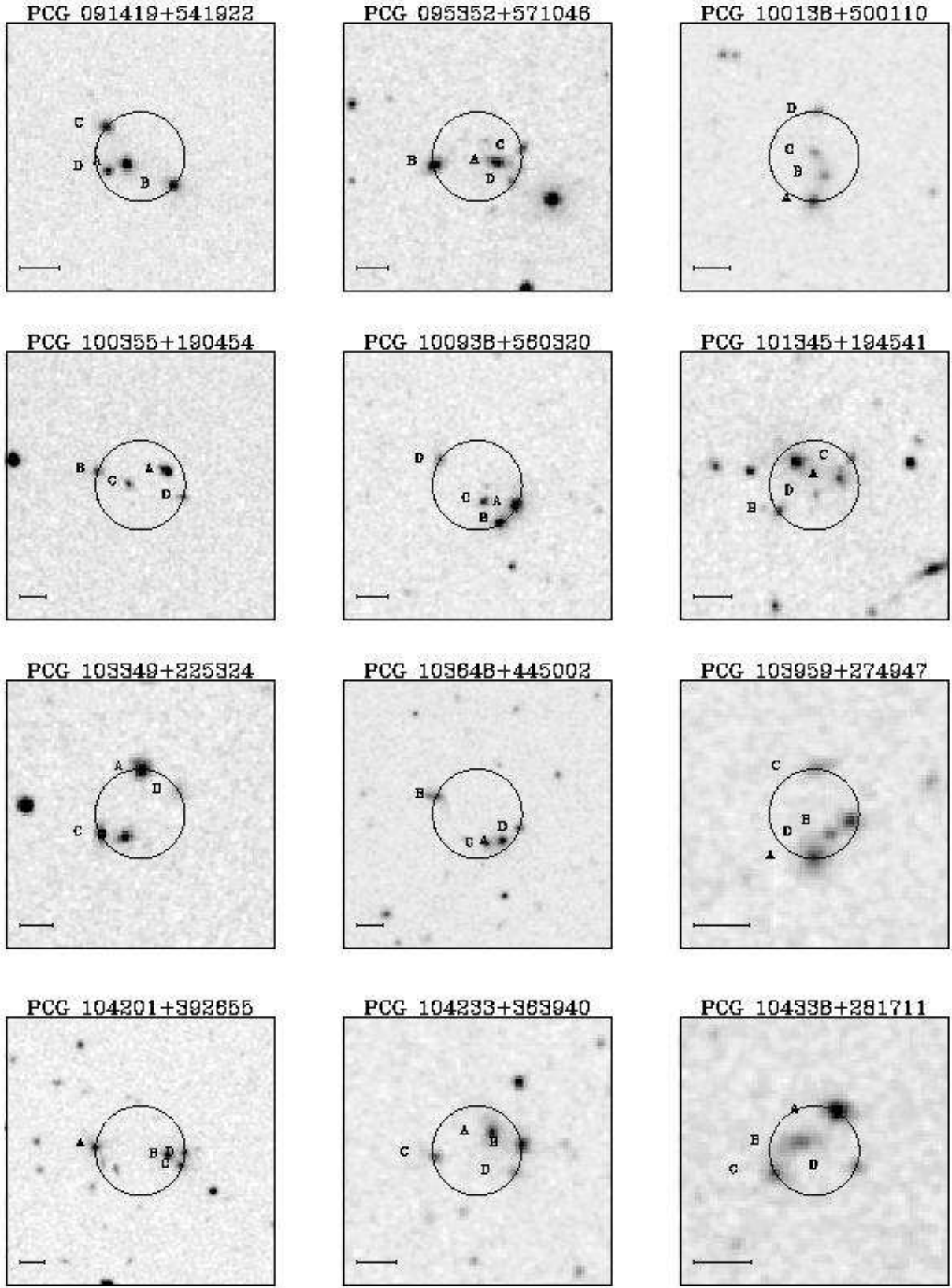}
\caption{DPOSS ($F$-plate) images of 
our 84 compact group candidates. The horizontal bar in the lower left
indicates a length of one arcminute.}
\label{Figure 2}
\end{figure}
\clearpage

\setcounter{figure}{1}
\begin{figure}
\plotone{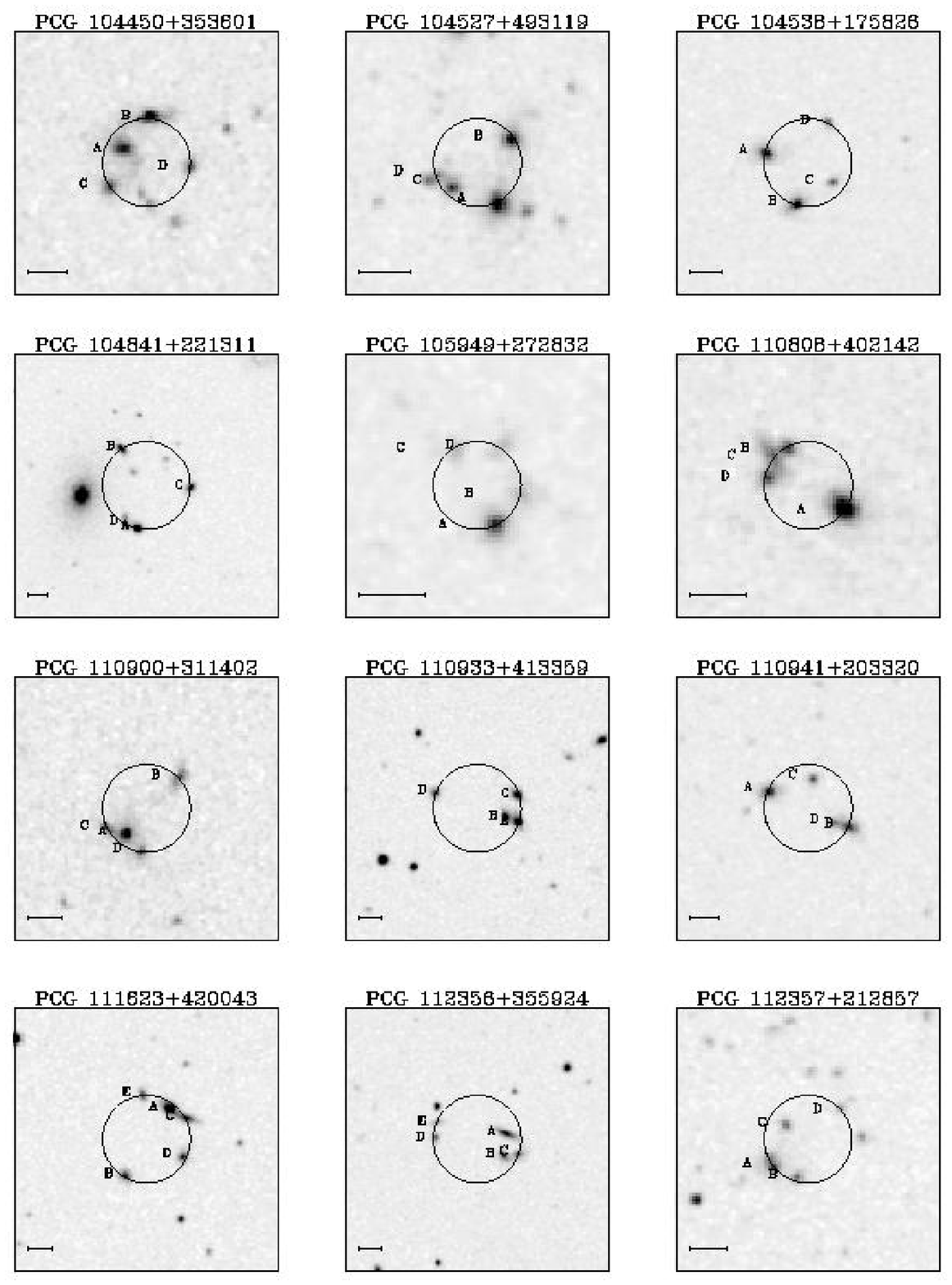} 
\caption{Continued}
\label{Figure 1}
\end{figure}
\clearpage

\setcounter{figure}{1}
\begin{figure}
\plotone{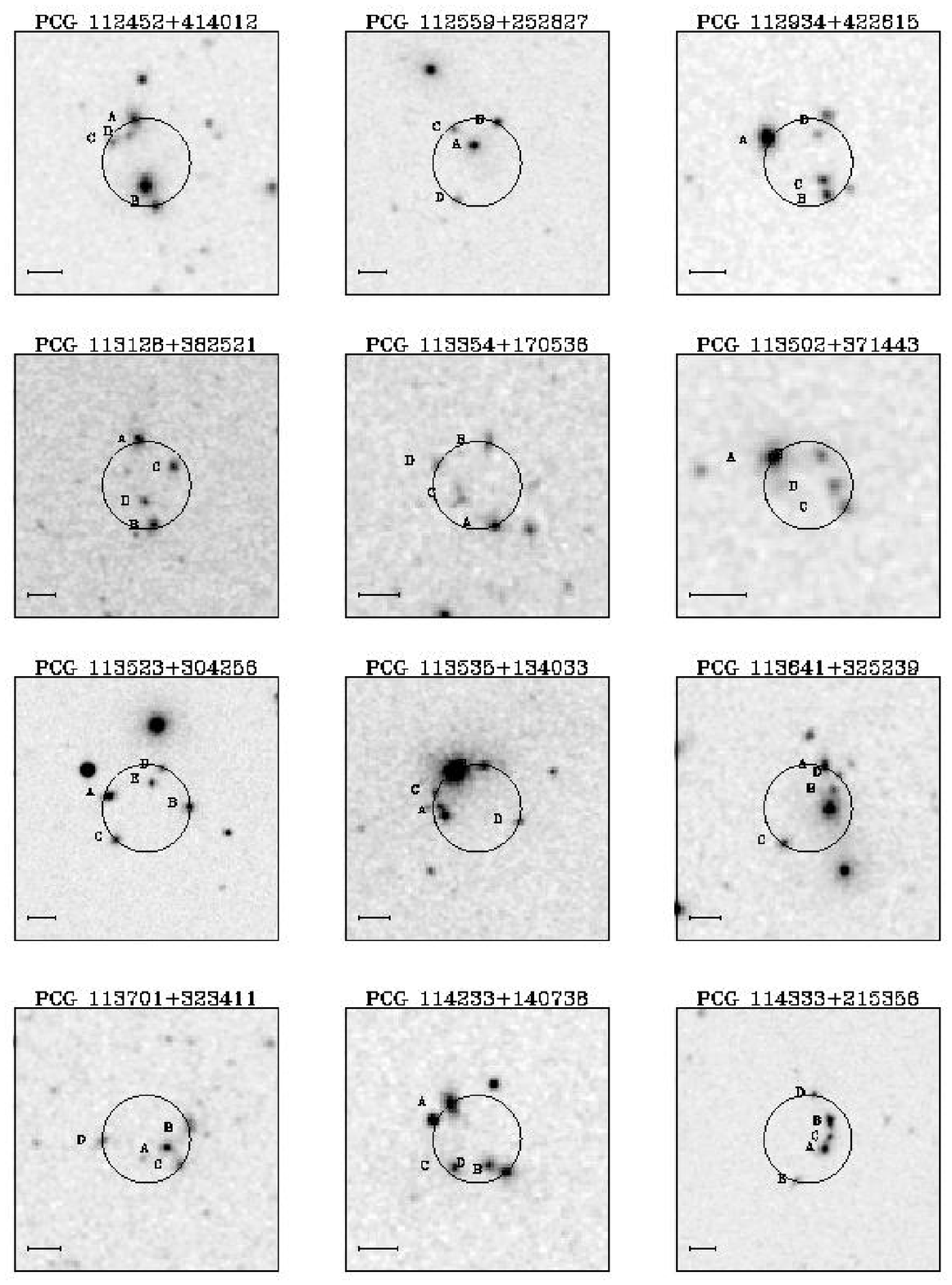}
\caption{Continued}
\label{Figure 2}
\end{figure}
\clearpage

\setcounter{figure}{1}
\begin{figure}
\plotone{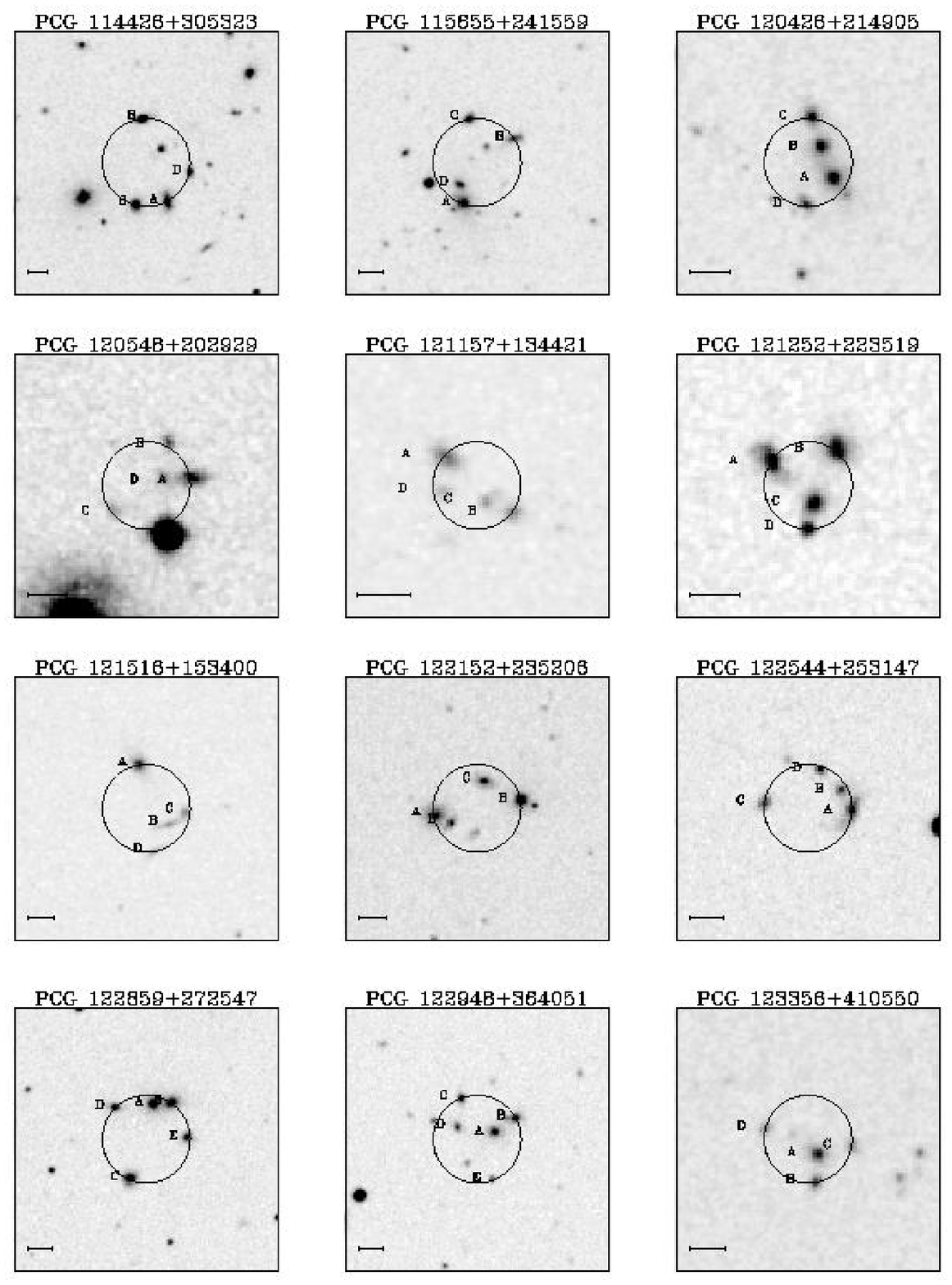}
\caption{Continued}
\label{Figure 3}
\end{figure}
\clearpage

\setcounter{figure}{1}
\begin{figure}
\plotone{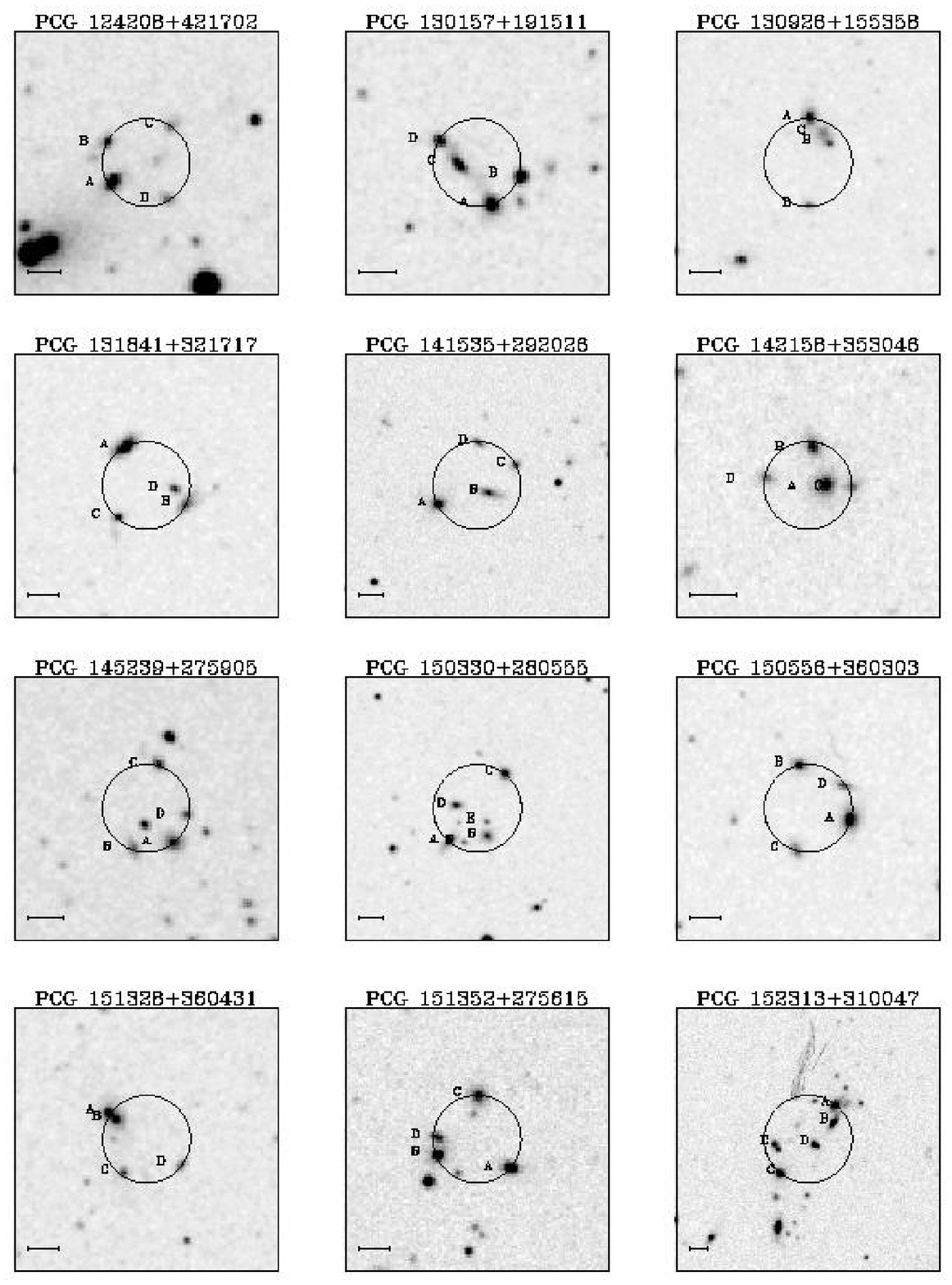}
\caption{Continued}
\label{Figure 4}
\end{figure}
\clearpage

\setcounter{figure}{1}
\begin{figure}
\plotone{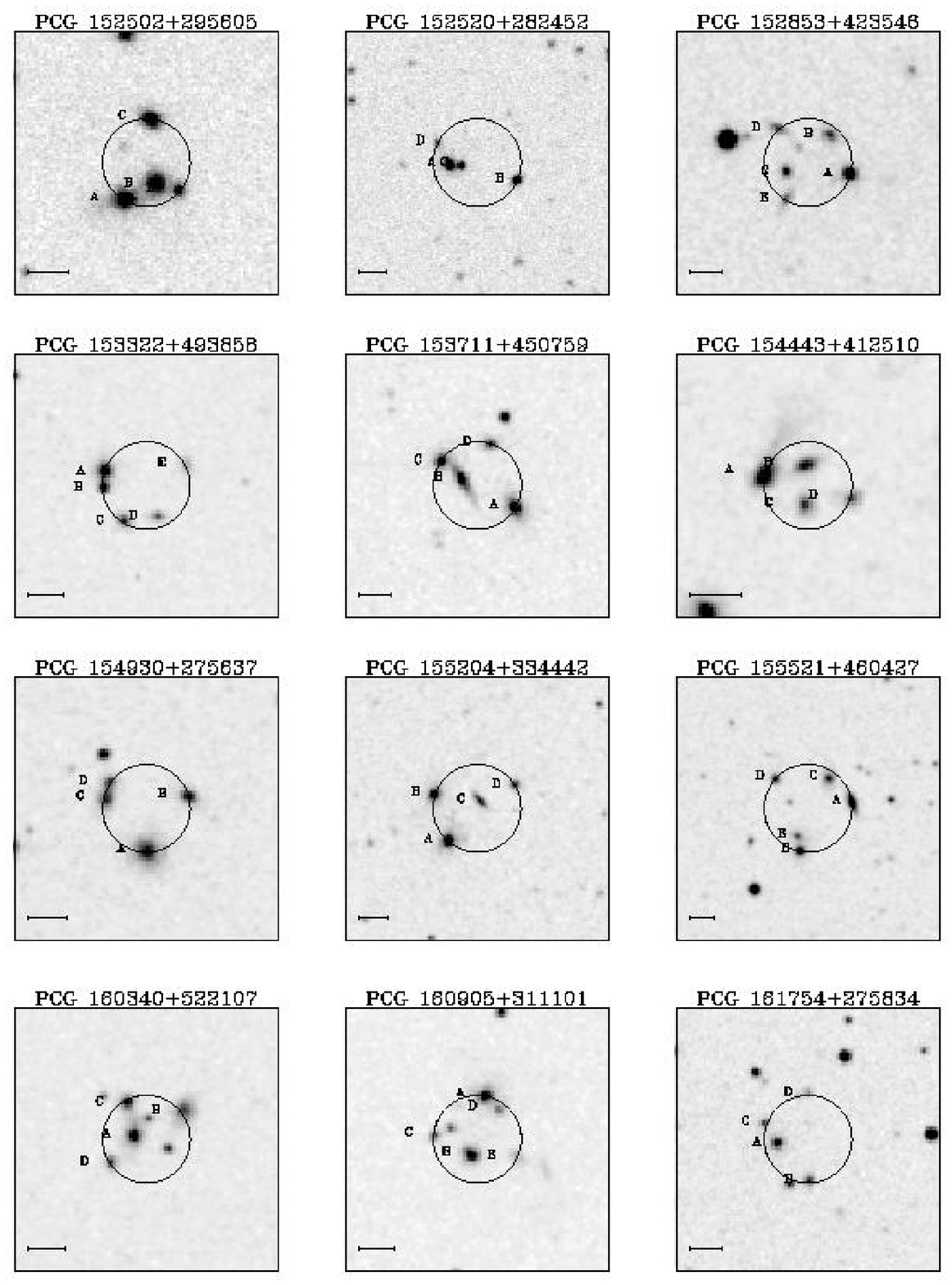}
\caption{Continued}
\label{Figure 5}
\end{figure}
\clearpage

\setcounter{figure}{1}
\begin{figure}
\plotone{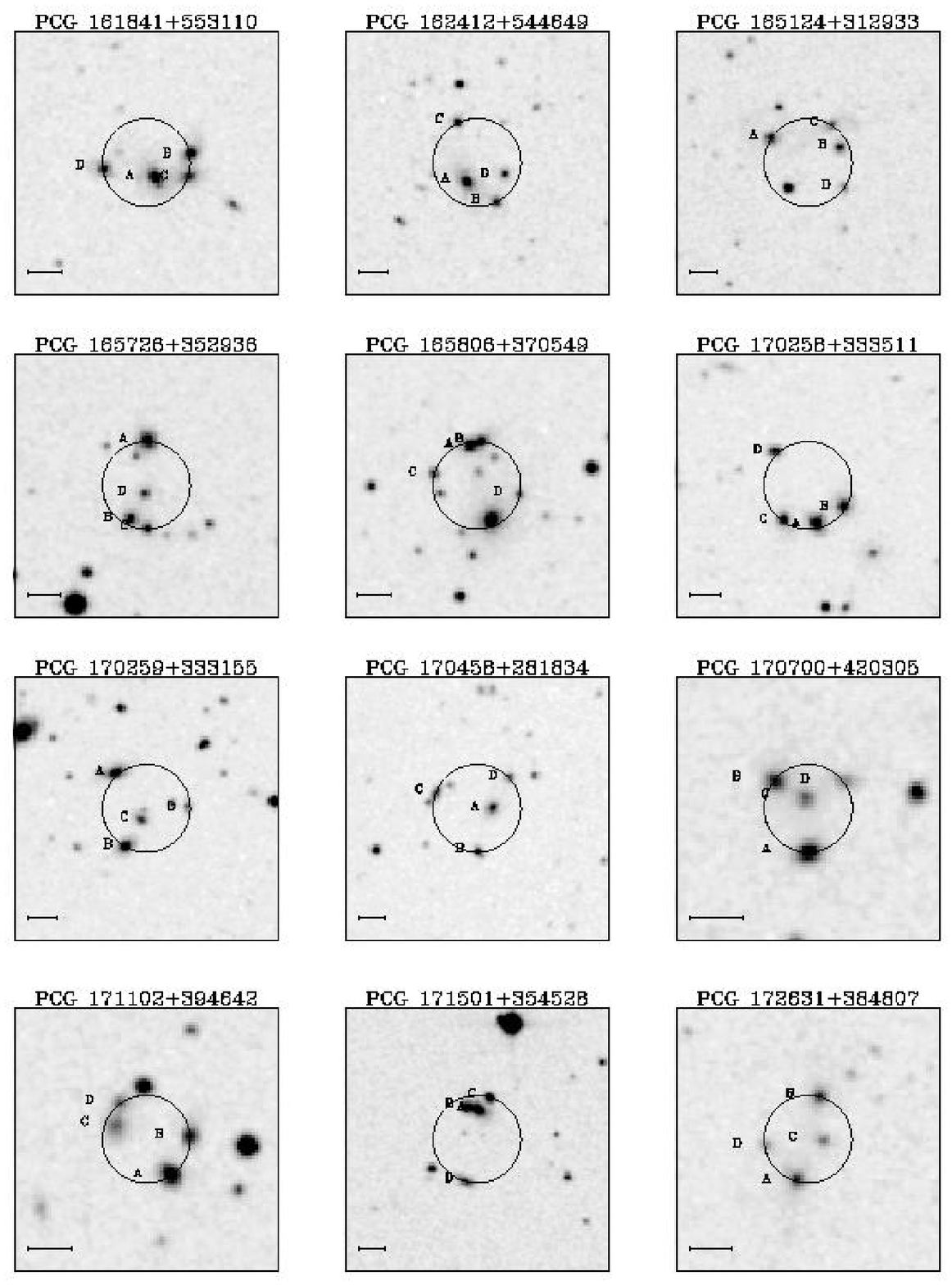}
\caption{Continued}
\label{Figure 6}
\end{figure}
\clearpage

\begin{figure}
\plotone{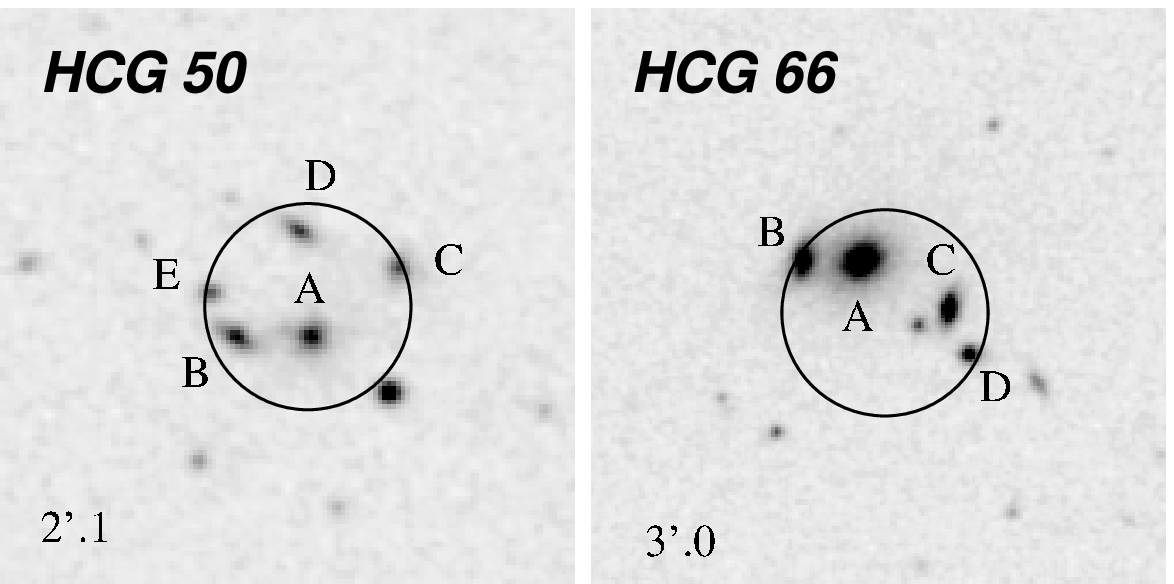}  
\caption{Two Hickson groups, HCG 50 and 66, which are at redshifts 0.14 and 0.07,
respectively. The images have been extracted from the same digitized
plate material used in our search. They show striking similarities to
our sample, suggesting our efficiency in extending the search of
Hickson-like objects to this largely unexplored redshift range. As in
Figure 2, the size of the field is indicated on the lower left.}
\label{Figure 3}
\end{figure}

\begin{deluxetable}{cccccccccccc}
\tabletypesize{\scriptsize}
\tablecaption{Characteristic Parameters of the Groups in our Sample}
\tablewidth{0pt}
\tablehead{
\colhead{Name} &\colhead{RA} &\colhead{DEC}& \colhead{Rad} & \colhead{Mag}
& \colhead{$<\mu>$}& \colhead{$\Delta \rm mag_{\rm comp}$}& \colhead{$\Delta \rm mag_{\rm isol}$}&Number \\
\colhead{}&  \colhead{(J2000)}&  \colhead{(J2000)}     &  \colhead{arcmin}      & \colhead{mag}
&\colhead{$(\rm mag arcsec^{-2})$}&\colhead{mag}&\colhead{mag}&
}
\startdata
PCG0914+5419& 09 14 19&+54 19 22 &0.27&15.76&23.09&1.10&3.00&4\\
PCG0953+5710& 09 53 52&+57 10 46 &0.35&15.50&23.36&1.86&2.80&4\\
PCG1001+5001& 10 01 38&+50 01 10 &0.31&16.14&23.70&1.54&2.19&4\\
PCG1003+1904& 10 03 55&+19 04 54 &0.41&15.60&23.80&1.49&3.00&4\\
PCG1009+5603& 10 09 38&+56 03 20 &0.36&15.38&23.30&1.53&3.00&4\\
PCG1013+1945& 10 13 45&+19 45 41 &0.29&15.96&23.42&1.98&2.85&4\\
PCG1033+2253& 10 33 49&+22 53 24 &0.33&15.46&23.18&1.67&3.00&4\\
PCG1036+4450& 10 36 48&+44 50 02 &0.43&15.45&23.75&1.22&1.74&4\\
PCG1039+2749& 10 39 59&+27 49 47 &0.20&16.09&22.71&1.77&3.00&4\\
PCG1042+3926& 10 42 01&+39 26 55 &0.46&15.14&23.58&0.58&1.50&4\\
PCG1042+3639& 10 42 33&+36 39 40 &0.28&15.74&23.07&1.89&2.86&4\\
PCG1043+2817& 10 43 38&+28 17 11 &0.19&16.04&22.60&1.98&3.00&4\\
PCG1044+3536& 10 44 50&+35 36 01 &0.28&15.12&22.51&1.79&2.32&4\\
PCG1045+4931& 10 45 27&+49 31 19 &0.22&15.59&22.38&1.70&2.54&4\\
PCG1045+1758& 10 45 38&+17 58 26 &0.34&15.58&23.35&1.53&3.00&4\\
PCG1048+2213& 10 48 41&+22 13 11 &0.56&14.88&23.77&0.70&2.01&4\\
PCG1059+2728& 10 59 49&+27 28 32 &0.16&16.41&22.62&1.94&3.00&4\\
PCG1108+4021& 11 08 08&+40 21 42 &0.19&16.03&22.57&1.71&3.00&4\\
PCG1109+3114& 11 09 00&+31 14 02 &0.32&15.33&23.00&1.61&2.46&4\\
PCG1109+4133& 11 09 33&+41 33 59 &0.49&15.30&23.88&0.96&2.16&4\\
PCG1109+2033& 11 09 41&+20 33 20 &0.37&15.88&23.85&1.62&2.77&4\\
PCG1116+4200& 11 16 23&+42 00 43 &0.44&15.19&23.55&1.72&3.00&5\\
PCG1123+3559& 11 23 56&+35 59 24 &0.47&15.18&23.67&1.95&3.00&5\\
PCG1123+2128& 11 23 57&+21 28 57 &0.29&15.72&23.20&1.79&2.88&4\\
PCG1124+4140& 11 24 52&+41 40 12 &0.32&16.09&23.77&1.95&2.53&4\\
PCG1125+2528& 11 25 59&+25 28 27 &0.40&15.38&23.53&1.84&3.00&4\\
PCG1129+4226& 11 29 34&+42 26 15 &0.30&15.56&23.07&1.81&2.99&4\\
PCG1131+3825& 11 31 28&+38 25 21 &0.39&15.79&23.87&1.33&2.71&4\\
PCG1133+1705& 11 33 54&+17 05 36 &0.27&16.26&23.54&1.56&2.68&4\\
PCG1135+3714& 11 35 02&+37 14 43 &0.19&15.90&22.49&1.95&2.55&4\\
PCG1135+3042& 11 35 23&+30 42 56 &0.40&15.49&23.64&1.25&3.00&5\\
PCG1135+1340& 11 35 35&+13 40 33 &0.35&15.51&23.36&1.86&2.95&4\\
PCG1136+3252& 11 36 41&+32 52 39 &0.35&15.78&23.63&1.78&2.80&4\\
PCG1137+3234& 11 37 01&+32 34 11 &0.34&15.48&23.29&0.96&2.43&4\\
PCG1142+1407& 11 42 33&+14 07 38 &0.28&15.40&22.76&1.25&3.00&4\\
PCG1143+2153& 11 43 33&+21 53 56 &0.42&15.71&23.96&1.83&2.41&5\\
PCG1144+3053& 11 44 26&+30 53 23 &0.54&14.76&23.54&0.97&1.53&4\\
PCG1156+2415& 11 56 55&+24 15 59 &0.44&15.31&23.65&1.25&2.70&4\\
PCG1204+2149& 12 04 26&+21 49 05 &0.27&15.19&22.48&1.30&2.87&4\\
PCG1205+2029& 12 05 48&+20 29 29 &0.28&15.91&23.25&1.97&3.00&4\\
PCG1211+1344& 12 11 57&+13 44 21 &0.21&16.20&22.93&1.45&3.00&4\\
PCG1212+2235& 12 12 52&+22 35 19 &0.22&15.21&22.02&1.22&3.00&4\\
PCG1215+1534& 12 15 16&+15 34 00 &0.43&15.49&23.78&1.76&3.00&4\\
PCG1221+2352& 12 21 52&+23 52 06 &0.40&15.31&23.45&1.38&2.31&4\\
PCG1225+2531& 12 25 44&+25 31 47 &0.31&16.11&23.74&1.40&2.84&4\\
PCG1228+2725& 12 28 59&+27 25 47 &0.46&14.90&23.32&1.08&2.59&5\\
PCG1229+3640& 12 29 48&+36 40 51 &0.45&15.49&23.87&1.91&2.96&5\\
PCG1233+4105& 12 33 56&+41 05 50 &0.31&16.25&23.85&1.66&3.00&4\\
PCG1242+4217& 12 42 08&+42 17 02 &0.33&15.52&23.23&1.85&2.56&4\\
PCG1301+1915& 13 01 57&+19 15 11 &0.29&15.39&22.86&0.88&2.24&4\\
PCG1309+1553& 13 09 26&+15 53 58 &0.35&16.05&23.89&1.67&3.00&4\\
PCG1318+3217& 13 18 41&+32 17 17 &0.36&15.66&23.58&1.92&3.00&4\\
PCG1415+2920& 14 15 35&+29 20 26 &0.44&15.38&23.72&1.50&2.46&4\\
PCG1421+3530& 14 21 58&+35 30 46 &0.23&16.11&23.06&1.72&3.00&4\\
PCG1452+2759& 14 52 39&+27 59 05 &0.30&16.09&23.62&1.29&2.43&4\\
PCG1503+2805& 15 03 30&+28 05 55 &0.44&15.37&23.70&1.97&2.75&6\\
PCG1505+3603& 15 05 56&+36 03 03 &0.35&15.75&23.63&1.55&3.00&4\\
PCG1513+3604& 15 13 28&+36 04 31 &0.35&16.03&23.91&1.46&2.57&4\\
PCG1513+2756& 15 13 52&+27 56 15 &0.35&15.50&23.33&0.94&2.50&4\\
PCG1523+3100& 15 23 13&+31 00 47 &0.61&14.71&23.77&0.87&2.16&5\\
PCG1525+2956& 15 25 02&+29 56 05 &0.27&15.07&22.33&1.42&3.00&4\\
PCG1525+2824& 15 25 20&+28 24 52 &0.41&15.78&23.96&1.82&2.71&4\\
PCG1528+4235& 15 28 53&+42 35 46 &0.34&15.84&23.63&1.71&3.00&5\\
PCG1533+4938& 15 33 22&+49 38 58 &0.31&15.73&23.33&1.66&3.00&5\\
PCG1537+4507& 15 37 11&+45 07 59 &0.34&15.31&23.10&1.35&3.00&4\\
PCG1544+4125& 15 44 43&+41 25 10 &0.21&16.27&23.05&1.79&3.00&4\\
PCG1549+2756& 15 49 30&+27 56 37 &0.28&15.70&23.09&1.67&3.00&4\\
PCG1552+3344& 15 52 04&+33 44 42 &0.37&15.67&23.64&1.61&3.00&4\\
PCG1555+4604& 15 55 21&+46 04 27 &0.46&15.45&23.88&1.52&2.23&5\\
PCG1603+5221& 16 03 40&+52 21 07 &0.29&15.77&23.20&1.48&2.27&4\\
PCG1609+3111& 16 09 05&+31 11 01 &0.30&15.64&23.17&1.95&2.81&5\\
PCG1617+2758& 16 17 54&+27 58 34 &0.34&15.99&23.81&1.64&2.36&4\\
PCG1618+5531& 16 18 41&+55 31 10 &0.32&15.30&23.00&1.02&2.05&4\\
PCG1624+5446& 16 24 12&+54 46 49 &0.36&15.68&23.62&1.72&3.00&4\\
PCG1651+3129& 16 51 24&+31 29 33 &0.40&15.65&23.79&1.71&2.86&4\\
PCG1657+3529& 16 57 26&+35 29 36 &0.33&15.66&23.38&1.38&2.75&4\\
PCG1658+3705& 16 58 06&+37 05 49 &0.34&15.77&23.56&1.66&2.42&4\\
PCG1702+3335& 17 02 58&+33 35 11 &0.35&15.43&23.27&1.01&3.00&4\\
PCG1702+3331& 17 02 59&+33 31 55 &0.38&15.61&23.62&1.92&2.52&4\\
PCG1704+2818& 17 04 57&+28 18 34 &0.41&15.62&23.84&0.97&1.71&4\\
PCG1707+4203& 17 07 00&+42 03 05 &0.21&15.77&22.49&1.63&3.00&4\\
PCG1711+3946& 17 11 02&+39 46 42 &0.25&15.70&22.80&1.53&2.03&4\\
PCG1715+3545& 17 15 01&+35 45 28 &0.42&15.19&23.42&1.15&3.00&4\\
PCG1726+3848& 17 26 31&+38 48 07 &0.26&16.11&23.28&1.94&2.66&4\\
\enddata
\end{deluxetable}
\begin{deluxetable}{cccccccccccc}
\tabletypesize{\scriptsize}
\tablecaption{Catalog of Groups}
\tablewidth{0pt}
\tablehead{
\colhead{Name} &\colhead{RA} &\colhead{DEC}& \colhead{r} & \colhead{g-r}
& \colhead{PA}& \colhead{Ellip.} \\
\colhead{}&  \colhead{(J2000)}&  \colhead{(J2000)}     &  \colhead{mag}      & \colhead{mag}
&\colhead{$(\circ)$}&\colhead{}
}
\startdata
PCG 0914+5419&  &           &        &               &            &   &   &  \\
A&09 14 20.50&+54 19 20.21&16.88& 0.30& 34&0.05\\
B&09 14 18.52&+54 19 11.79&17.11& 0.31& 04&0.08\\
C&09 14 21.28&+54 19 34.04&17.38& 0.26& 40&0.30\\
D&09 14 21.31&+54 19 18.26&17.98& 0.28&-34&0.08\\
PCG 0953+5710&  &           &        &               &            &   &   &  \\
A&09 53 51.81&+57 10 43.57&16.40& 0.27& 09&0.48\\
B&09 53 55.46&+57 10 42.67&16.53& 0.30&-23&0.31\\
C&09 53 50.35&+57 10 50.01&17.97& 0.26&-49&0.34\\
D&09 53 50.96&+57 10 33.99&18.26& 0.18& 85&0.25\\
PCG 1001+5001&  &           &        &               &            &   &   &  \\
A&10 01 38.86&+50 00 52.71&16.85& 0.61&-65&0.11\\
B&10 01 38.41&+50 01  3.36&17.81& 0.37&-72&0.08\\
C&10 01 38.75&+50 01 12.21&18.25& 0.46& 39&0.48\\
D&10 01 38.65&+50 01 29.28&18.39& 0.31&-37&0.19\\
PCG 1003+1904&  &           &        &               &            &   &   &  \\
A&10 03 54.22&+19 05  2.11&16.37& 0.38& 33&0.27\\
B&10 03 56.92&+19 05  2.11&17.23& 0.34& 48&0.07\\
C&10 03 55.69&+19 04 55.09&17.56& 0.36& 56&0.31\\
D&10 03 53.61&+19 04 47.20&17.86& 0.29& 12&0.26\\
PCG 1009+5603&  &           &        &               &            &   &   &  \\
A&10 09 36.39&+56 03 11.66&16.23& 0.43&-70&0.24\\
B&10 09 37.21&+56 03  3.46&16.62& 0.35&-33&0.43\\
C&10 09 38.21&+56 03 12.92&17.75& 0.33&-34&0.16\\
D&10 09 40.91&+56 03 32.43&17.77& 0.30&-83&0.42\\
 &           &            &     &     &   &    \\
PCG 1013+1945&  &           &        &               &            &   &   &  \\
A&10 13 45.16&+19 45 44.28&16.67& 0.60& 87&0.12\\
B&10 13 46.87&+19 45 30.71&17.60& 0.43&-55&0.37\\
C&10 13 44.87&+19 45 51.52&17.82& 0.48&-56&0.27\\
D&10 13 45.84&+19 45 37.61&18.65& 0.48& 78&0.25\\
PCG 1033+2253&  &           &        &               &            &   &   &  \\
A&10 33 48.96&+22 53 44.60&16.37&-0.07& 74&0.13\\
B&10 33 49.50&+22 53 14.43&16.91& 0.43&-21&0.27\\
C&10 33 50.26&+22 53 15.68&17.15& 0.52& 83&0.46\\
D&10 33 47.75&+22 53 34.15&18.03& 0.06& 69&0.13\\
PCG 1036+4450&  &           &        &               &            &   &   &  \\
A&10 36 46.97&+44 49 45.12&16.44& 0.40& 00&0.17\\
B&10 36 50.49&+44 50 11.76&16.74& 0.12& 34&0.37\\
C&10 36 47.76&+44 49 43.75&17.37& 0.48& 20&0.16\\
D&10 36 46.00&+44 49 52.61&17.67& 0.32&-41&0.23\\
PCG 1039+2749&  &           &        &               &            &   &   &  \\
A&10 39 59.23&+27 49 35.72&16.86& 0.07&-56&0.06\\
B&10 39 58.52&+27 49 44.98&17.46& 0.48&-05&0.11\\
C&10 39 59.13&+27 49 59.48&18.33& 0.22&-14&0.47\\
D&10 39 58.90&+27 49 41.88&18.63& 0.57&-17&0.35\\
PCG 1042+3926&  &           &        &               &            &   &   &  \\
A&10 42  3.74&+39 26 57.98&16.43& 0.41& 48&0.15\\
B&10 41 59.88&+39 26 51.60&16.45& 0.47& 25&0.28\\
C&10 41 59.25&+39 26 45.02&16.76& 0.48& 44&0.21\\
D&10 41 59.02&+39 26 52.37&17.01& 0.39& 59&0.20\\
PCG 1042+3639&  &           &        &               &            &   &   &  \\
A&10 42 33.35&+36 39 46.48&16.65& 0.19& 77&0.26\\
B&10 42 32.47&+36 39 41.84&16.85& 0.08& 82&0.26\\
C&10 42 35.19&+36 39 38.38&17.93& 0.56& 25&0.45\\
D&10 42 32.71&+36 39 31.78&18.54& 0.25&-25&0.13\\
PCG 1043+2817&  &           &        &               &            &   &   &  \\
A&10 43 38.19&+28 17 21.23&16.87&-0.06& 14&0.27\\
B&10 43 38.97&+28 17 13.09&17.44&-0.05&-13&0.56\\
C&10 43 39.40&+28 17  5.49&17.84& 0.91&-29&0.02\\
D&10 43 37.83&+28 17  6.68&18.84& 0.58&-30&0.08\\
 &           &            &     &     &   &    \\
PCG 1044+3536&  &           &        &               &            &   &   &  \\
A&10 44 50.95&+35 36  6.28&16.13&-0.21& 45&0.35\\
B&10 44 50.05&+35 36 18.41&16.16& 0.18&-02&0.53\\
C&10 44 51.40&+35 35 51.99&17.19& 0.42&-58&0.10\\
D&10 44 48.87&+35 35 59.21&17.91& 0.22&-62&0.22\\
PCG 1045+4931&  &           &        &               &            &   &   &  \\
A&10 45 26.68&+49 31  8.25&16.37& 0.57&-83&0.22\\
B&10 45 26.20&+49 31 26.43&16.92& 0.42&-70&0.18\\
C&10 45 28.04&+49 31 13.43&17.88& 0.48& 42&0.05\\
D&10 45 28.59&+49 31 16.11&18.07& 0.74&-18&0.54\\
PCG 1045+1758&  &           &        &               &            &   &   &  \\
A&10 45 39.90&+17 58 31.48&16.58& 0.28& 02&0.34\\
B&10 45 38.98&+17 58  7.86&16.59& 0.34&-32&0.24\\
C&10 45 37.78&+17 58 17.54&17.92& 0.29&-37&0.33\\
D&10 45 37.90&+17 58 45.16&18.11& 0.35& 48&0.30\\
PCG 1048+2213&  &           &        &               &            &   &   &  \\
A&10 48 42.52&+22 12 38.41&16.13& 0.30& 44&0.22\\
B&10 48 43.35&+22 13 39.50&16.29& 0.17& 43&0.36\\
C&10 48 39.54&+22 13  9.05&16.37& 0.32&-57&0.22\\
D&10 48 43.14&+22 12 41.72&16.83& 0.18& 80&0.49\\
PCG 1059+2728&  &           &        &               &            &   &   &  \\
A&10 59 49.74&+27 28 23.02&16.89& 0.40&-74&0.10\\
B&10 59 49.29&+27 28 29.75&18.62& 0.73&-55&0.21\\
C&10 59 50.44&+27 28 39.97&18.71& 0.97& 07&0.23\\
D&10 59 49.61&+27 28 40.77&18.83& 0.70&-49&0.38\\
PCG 1108+4021&  &           &        &               &            &   &   &  \\
A&11 08  7.25&+40 21 35.08&16.64& 0.26& 35&0.19\\
B&11 08  8.53&+40 21 51.08&17.87& 1.28&-26&0.30\\
C&11 08  8.84&+40 21 49.10&18.25& 0.02& 57&0.51\\
D&11 08  8.99&+40 21 43.64&18.34& 0.26&-42&0.18\\
 &           &            &     &     &   &    \\
PCG 1109+3114&  &           &        &               &            &   &   &  \\
A&11 09  1.09&+31 13 51.24&16.06& 0.42&-36&0.12\\
B&11 08 59.28&+31 14 15.86&16.95& 0.18&-64&0.48\\
C&11 09  1.70&+31 13 53.44&17.33& 0.71& 38&0.61\\
D&11 09  0.58&+31 13 43.54&17.67& 0.29&-72&0.27\\
PCG 1109+4133&  &           &        &               &            &   &   &  \\
A&11 09 31.06&+41 33 49.28&16.43& 0.35& 40&0.23\\
B&11 09 31.75&+41 33 52.42&16.76& 0.30& 66&0.26\\
C&11 09 31.05&+41 34  7.18&16.82& 0.33& 36&0.32\\
D&11 09 35.97&+41 34  9.62&17.40& 0.39&-18&0.03\\
PCG 1109+2033&  &           &        &               &            &   &   &  \\
A&11 09 42.65&+20 33 29.67&16.88& 0.11&-04&0.28\\
B&11 09 39.79&+20 33 11.20&16.90& 0.38& 21&0.40\\
C&11 09 41.07&+20 33 35.53&18.11& 0.35&-45&0.08\\
D&11 09 40.29&+20 33 13.28&18.50& 0.55&-30&0.05\\
PCG 1116+4200&  &           &        &               &            &   &   &  \\
A&11 16 21.83&+42 01  1.99&16.12& 0.08& 41&0.39\\
B&11 16 24.28&+42 00 21.13&17.03&-0.06& 64&0.20\\
C&11 16 20.95&+42 00 56.02&17.14& 0.40& 18&0.57\\
D&11 16 21.14&+42 00 33.01&17.31& 0.50&-58&0.22\\
E&11 16 23.31&+42 01 10.27&17.84& 0.26& 77&0.22\\
PCG 1123+3559&  &           &        &               &            &   &   &  \\
A&11 23 55.11&+35 59 28.14&16.09& 0.03& 24&0.66\\
B&11 23 55.17&+35 59 13.88&16.76& 0.01& 43&0.39\\
C&11 23 54.45&+35 59 15.00&17.28& 0.30&-07&0.28\\
D&11 23 58.86&+35 59 24.36&17.44& 0.27& 68&0.06\\
E&11 23 58.80&+35 59 34.73&18.05& 0.29&-13&0.48\\
 &           &            &     &     &   &    \\
 &           &            &     &     &   &    \\
 &           &            &     &     &   &    \\
 &           &            &     &     &   &    \\
PCG 1123+2128&  &           &        &               &            &   &   &  \\
A&11 23 58.82&+21 28 46.84&16.33& 0.66& 57&0.20\\
B&11 23 58.09&+21 28 41.96&17.60& 0.54&-46&0.14\\
C&11 23 58.40&+21 29  3.05&17.83& 0.56&-80&0.18\\
D&11 23 56.82&+21 29  8.38&18.11& 0.60&-39&0.33\\
PCG 1124+4140&  &           &        &               &            &   &   &  \\
A&11 24 52.78&+41 40 31.37&16.75& 0.76& 62&0.20\\
B&11 24 51.90&+41 39 54.03&17.49& 0.61& 44&0.24\\
C&11 24 53.61&+41 40 21.61&18.69& 0.18&-37&0.35\\
D&11 24 52.95&+41 40 24.83&18.70& 0.57&-44&0.51\\
PCG 1125+2528&  &           &        &               &            &   &   &  \\
A&11 25 59.90&+25 28 36.26&16.07& 0.50&-16&0.13\\
B&11 25 58.97&+25 28 49.15&17.02& 0.55& 05&0.09\\
C&11 26  0.71&+25 28 45.20&17.39& 0.54& 16&0.17\\
D&11 26  0.59&+25 28  6.35&17.91& 0.09& 56&0.00\\
PCG 1129+4226&  &           &        &               &            &   &   &  \\
A&11 29 36.13&+42 26 23.21&16.08&-0.09&-81&0.28\\
B&11 29 34.00&+42 25 59.08&17.69& 0.13&-85&0.11\\
C&11 29 34.11&+42 26  4.85&17.80& 0.64& 29&0.20\\
D&11 29 33.90&+42 26 31.14&17.89& 0.10&-11&0.02\\
PCG 1131+3825&  &           &        &               &            &   &   &  \\
A&11 31 28.78&+38 25 44.51&16.70& 0.63&-75&0.20\\
B&11 31 28.23&+38 24 58.25&17.20& 0.49&-73&0.25\\
C&11 31 27.22&+38 25 29.02&17.70& 0.19&-52&0.06\\
D&11 31 28.62&+38 25 11.03&18.02& 0.34& 47&0.04\\
PCG 1133+1705&  &           &        &               &            &   &   &  \\
A&11 33 54.15&+17 05 22.02&16.95& 0.46& 41&0.15\\
B&11 33 54.31&+17 05 52.29&18.01& 0.34&-81&0.48\\
C&11 33 55.05&+17 05 32.68&18.30& 0.58& 71&0.39\\
D&11 33 55.61&+17 05 44.34&18.51& 0.47&-75&0.15\\
 &           &            &     &     &   &    \\
PCG 1135+3714&  &           &        &               &            &   &   &  \\
A&11 35  3.23&+37 14 49.76&16.37& 0.27&-65&0.13\\
B&11 35  2.16&+37 14 50.28&18.09& 0.42& 19&0.19\\
C&11 35  1.62&+37 14 36.37&18.26& 0.32& 47&0.12\\
D&11 35  1.85&+37 14 41.93&18.33& 0.23& 78&0.05\\
PCG 1135+3042&  &           &        &               &            &   &   &  \\
A&11 35 24.99&+30 43  3.18&16.68& 0.16&-23&0.34\\
B&11 35 21.51&+30 42 57.38&16.80& 0.38&-83&0.12\\
C&11 35 24.66&+30 42 38.74&17.54& 0.36&-52&0.12\\
D&11 35 22.71&+30 43 18.69&17.87& 0.20&-12&0.14\\
E&11 35 23.10&+30 43 10.49&17.93& 0.26& 42&0.09\\
PCG 1135+1340&  &           &        &               &            &   &   &  \\
A&11 35 36.68&+13 40 31.51&16.25& 0.30& 64&0.56\\
B&11 35 35.37&+13 40 53.51&16.91& 0.44& 00&0.40\\
C&11 35 36.91&+13 40 40.84&17.74& 0.14&-83&0.21\\
D&11 35 34.20&+13 40 26.94&18.11& 0.30&-22&0.27\\
PCG 1136+3252&  &           &        &               &            &   &   &  \\
A&11 36 40.57&+32 52 58.77&16.48& 0.08& 83&0.40\\
B&11 36 40.26&+32 52 46.82&17.56& 0.34&-88&0.34\\
C&11 36 42.10&+32 52 22.16&17.64&-0.06&-60&0.14\\
D&11 36 40.02&+32 52 54.51&18.26& 0.30& 87&0.15\\
PCG 1137+3234&  &           &        &               &            &   &   &  \\
A&11 37  1.00&+32 34  5.45&16.62& 0.62& 08&0.06\\
B&11 37  0.12&+32 34 14.77&16.73& 0.55&-88&0.38\\
C&11 37  0.51&+32 33 57.39&17.29& 0.68& 62&0.07\\
D&11 37  3.34&+32 34  9.20&17.58& 0.52&-33&0.34\\
PCG 1142+1407&  &           &        &               &            &   &   &  \\
A&11 42 33.85&+14 07 51.56&16.34& 0.28& 72&0.39\\
B&11 42 32.39&+14 07 25.61&16.74& 0.36& 12&0.14\\
C&11 42 33.78&+14 07 27.09&17.38& 0.30& 42&0.23\\
D&11 42 32.83&+14 07 28.27&17.59& 0.35& 44&0.08\\
PCG 1143+2153&  &           &        &               &            &   &   &  \\
A&11 43 33.15&+21 53 50.32&16.83&-0.01&-75&0.25\\
B&11 43 32.86&+21 54  5.44&16.85& 0.22& 79&0.35\\
C&11 43 32.93&+21 53 56.37&18.02& 0.26&-72&0.33\\
D&11 43 33.53&+21 54 21.46&18.09& 0.27& 22&0.03\\
E&11 43 34.29&+21 53 32.03&18.65& 0.24&-44&0.32\\
PCG 1144+3053&  &           &        &               &            &   &   &  \\
A&11 44 25.77&+30 52 55.12&16.01& 0.34& 87&0.34\\
B&11 44 27.07&+30 53 56.15&16.13& 0.33&-14&0.49\\
C&11 44 27.55&+30 52 52.93&16.17& 0.32&-70&0.03\\
D&11 44 24.46&+30 53 16.26&16.98& 0.33& 89&0.31\\
PCG 1156+2415&  &           &        &               &            &   &   &  \\
A&11 56 55.84&+24 15 35.10&16.07& 0.56& 80&0.21\\
B&11 56 53.53&+24 16 13.58&17.08& 0.34&-07&0.42\\
C&11 56 55.50&+24 16 25.89&17.29& 0.41&-22&0.31\\
D&11 56 55.95&+24 15 46.40&17.33& 0.52& 56&0.22\\
PCG 1204+2149&  &           &        &               &            &   &   &  \\
A&12 04 26.12&+21 48 59.65&16.18& 0.31& 55&0.24\\
B&12 04 26.44&+21 49 10.91&16.55& 0.35&-83&0.05\\
C&12 04 26.70&+21 49 22.00&16.98& 0.35& 73&0.18\\
D&12 04 26.85&+21 48 49.68&17.48& 0.29& 34&0.10\\
PCG 1205+2029&  &           &        &               &            &   &   &  \\
A&12 05 47.04&+20 29 30.37&16.46& 0.17& 03&0.53\\
B&12 05 47.66&+20 29 43.84&17.79&-0.11& 81&0.30\\
C&12 05 49.11&+20 29 18.46&18.15& 0.77&-21&0.24\\
D&12 05 47.81&+20 29 30.16&18.42& 0.08& 05&0.14\\
PCG 1211+1344&  &           &        &               &            &   &   &  \\
A&12 11 58.59&+13 44 29.76&16.96& 0.00& 33&0.37\\
B&12 11 57.29&+13 44 13.34&17.78& 0.45&-14&0.21\\
C&12 11 57.77&+13 44 16.77&18.34& 0.32&-33&0.45\\
D&12 11 58.65&+13 44 19.75&18.40& 0.41& 44&0.42\\
PCG 1212+2235&  &           &        &               &            &   &   &  \\
A&12 12 53.31&+22 35 26.56&16.37& 0.12& 48&0.40\\
B&12 12 51.93&+22 35 30.08&16.44& 0.02& 34&0.11\\
C&12 12 52.42&+22 35 14.28&16.86& 0.42&-59&0.16\\
D&12 12 52.55&+22 35  6.94&17.59& 0.39&-88&0.23\\
PCG 1215+1534&  &           &        &               &            &   &   &  \\
A&12 15 16.30&+15 34 25.29&16.12& 0.45& 65&0.28\\
B&12 15 15.07&+15 33 50.90&17.31& 0.53&-18&0.70\\
C&12 15 14.43&+15 33 57.85&17.62& 0.36&-65&0.40\\
D&12 15 15.71&+15 33 34.89&17.87& 0.03&-32&0.57\\
PCG 1221+2352&  &           &        &               &            &   &   &  \\
A&12 21 53.95&+23 52  2.68&16.39& 0.07&-09&0.12\\
B&12 21 50.50&+23 52 10.20&16.43& 0.46& 44&0.02\\
C&12 21 51.95&+23 52 20.96&17.24& 0.27& 14&0.41\\
D&12 21 53.33&+23 51 58.28&17.76& 0.59&-42&0.39\\
PCG 1225+2531&  &           &        &               &            &   &   &  \\
A&12 25 43.58&+25 31 45.16&16.83& 0.14&-69&0.54\\
B&12 25 43.88&+25 31 53.87&17.94& 0.27& 28&0.09\\
C&12 25 46.37&+25 31 48.87&18.10& 0.03&-37&0.23\\
D&12 25 44.57&+25 32  2.83&18.23& 0.37& 41&0.25\\
PCG 1228+2725&  &           &        &               &            &   &   &  \\
A&12 28 59.19&+27 26  8.81&16.19& 0.36&-18&0.13\\
B&12 28 58.37&+27 26  9.35&16.36& 0.30&-06&0.27\\
C&12 29  0.33&+27 25 21.83&16.62& 0.02&-14&0.07\\
D&12 29  1.02&+27 26  6.51&17.20& 0.34& 04&0.39\\
E&12 28 57.61&+27 25 47.32&17.27& 0.30& 07&0.07\\
PCG 1228+2725&  &           &        &               &            &   &   &  \\
A&12 28 58.37&+27 26  9.35&16.36& 0.30&-06&0.27\\
B&12 29  0.33&+27 25 21.83&16.62& 0.02&-14&0.07\\
C&12 29  1.02&+27 26  6.51&17.20& 0.34& 04&0.39\\
D&12 28 57.61&+27 25 47.32&17.27& 0.30& 07&0.07\\
PCG 1229+3640&  &           &        &               &            &   &   &  \\
A&12 29 47.52&+36 40 54.80&16.62& 0.52&-08&0.20\\
B&12 29 46.42&+36 41  4.12&16.79& 0.44&-11&0.43\\
C&12 29 49.31&+36 41 16.02&17.47& 0.47& 83&0.31\\
D&12 29 49.47&+36 40 58.66&17.72& 0.25& 51&0.28\\
E&12 29 47.66&+36 40 26.26&18.53& 0.58&-42&0.10\\
PCG 1233+4105&  &           &        &               &            &   &   &  \\
A&12 33 56.28&+41 05 43.67&16.96& 0.38& 37&0.17\\
B&12 33 56.36&+41 05 31.91&17.81& 0.56&-65&0.26\\
C&12 33 54.95&+41 05 46.69&18.42& 0.32& 64&0.18\\
D&12 33 58.18&+41 05 54.53&18.63& 0.29&-50&0.18\\
PCG 1242+4217&  &           &        &               &            &   &   &  \\
A&12 42  9.53&+42 16 52.46&16.09& 0.43&-47&0.37\\
B&12 42  9.78&+42 17 10.75&17.29& 0.50&-42&0.16\\
C&12 42  7.15&+42 17 18.74&17.90& 0.39&-45&0.20\\
D&12 42  7.30&+42 16 45.26&17.94& 0.32& 06&0.28\\
PCG 1301+1915&  &           &        &               &            &   &   &  \\
A&13 01 56.80&+19 14 54.93&16.57& 0.11& 74&0.16\\
B&13 01 55.98&+19 15  6.26&16.84& 0.17&-55&0.11\\
C&13 01 57.72&+19 15 11.27&16.88&-0.09& 55&0.52\\
D&13 01 58.24&+19 15 20.38&17.46& 0.22& 34&0.17\\
PCG 1309+1553&  &           &        &               &            &   &   &  \\
A&13 09 26.89&+15 54 19.48&16.73& 0.38& 85&0.24\\
B&13 09 26.26&+15 54  7.67&17.80& 0.15& 68&0.38\\
C&13 09 26.44&+15 54 12.20&18.05& 0.00& 20&0.40\\
D&13 09 26.91&+15 53 37.82&18.41& 0.47& 05&0.16\\
PCG 1318+3217&  &           &        &               &            &   &   &  \\
A&13 18 42.84&+32 17 36.82&16.27& 0.11&-37&0.44\\
B&13 18 40.47&+32 17  8.49&17.37&-0.05&-47&0.51\\
C&13 18 43.17&+32 17  2.22&17.92& 0.01&-25&0.26\\
D&13 18 40.94&+32 17 15.50&18.18& 0.45& 34&0.21\\
PCG 1415+2920&  &           &        &               &            &   &   &  \\
A&14 15 36.91&+29 20 14.78&16.12& 0.29&-21&0.19\\
B&14 15 34.57&+29 20 21.98&16.96& 0.46& 12&0.64\\
C&14 15 33.33&+29 20 38.30&17.55& 0.11&-61&0.13\\
D&14 15 35.05&+29 20 51.51&17.62& 0.15& 31&0.44\\
PCG 1421+3530&  &           &        &               &            &   &   &  \\
A&14 21 58.34&+35 30 45.36&16.88& 0.51&-41&0.04\\
B&14 21 58.66&+35 30 57.64&17.40& 0.56& 63&0.31\\
C&14 21 57.67&+35 30 45.24&18.50& 0.53&-18&0.23\\
D&14 21 59.93&+35 30 47.61&18.61& 0.42& 01&0.20\\
PCG 1452+2759&  &           &        &               &            &   &   &  \\
A&14 52 39.07&+27 58 50.41&16.92& 0.09& 10&0.13\\
B&14 52 40.31&+27 58 48.18&17.79& 0.52& 69&0.38\\
C&14 52 39.48&+27 59 22.67&17.90& 0.36& 49&0.27\\
D&14 52 38.65&+27 59  1.68&18.22& 0.54&-29&0.14\\
PCG 1503+2805&  &           &        &               &            &   &   &  \\
A&15 03 32.21&+28 05 34.80&16.66& 0.42&-44&0.53\\
B&15 03 30.51&+28 05 38.33&16.97& 0.30&-04&0.01\\
C&15 03 29.74&+28 06 15.66&17.08& 0.47& 62&0.25\\
D&15 03 31.90&+28 05 56.44&17.39& 0.23& 04&0.34\\
PCG 1505+3603&  &           &        &               &            &   &   &  \\
A&15 05 54.90&+36 02 57.44&16.46& 0.23&-80&0.26\\
B&15 05 56.92&+36 03 24.18&17.43& 0.23& 06&0.26\\
C&15 05 57.08&+36 02 42.97&17.84& 0.14& 66&0.28\\
D&15 05 55.20&+36 03 13.66&18.01& 0.15& 17&0.53\\
PCG 1513+3604&  &           &        &               &            &   &   &  \\
A&15 13 29.87&+36 04 43.68&16.91& 0.39& 11&0.25\\
B&15 13 29.63&+36 04 40.62&17.27& 0.48& 47&0.20\\
C&15 13 29.30&+36 04 14.44&18.29& 0.60&-36&0.02\\
D&15 13 27.05&+36 04 18.37&18.36& 0.24&-54&0.35\\
 &           &            &     &     &   &    \\
PCG 1513+2756&  &           &        &               &            &   &   &  \\
A&15 13 51.38&+27 56  1.32&16.68& 0.44& 09&0.24\\
B&15 13 53.99&+27 56  8.02&16.84& 0.59&-71&0.13\\
C&15 13 52.49&+27 56 36.24&17.06& 0.19& 72&0.14\\
D&15 13 53.97&+27 56 16.15&17.62&-0.06& 19&0.34\\
PCG 1523+3100&  &           &        &               &            &   &   &  \\
A&15 23 11.29&+31 01 15.92&16.07& 0.25&-45&0.25\\
B&15 23 11.37&+31 01  1.63&16.37& 0.53&-59&0.58\\
C&15 23 14.85&+31 00 18.65&16.38& 0.36& 50&0.10\\
D&15 23 12.63&+31 00 43.24&16.75& 0.39& 50&0.15\\
E&15 23 15.25&+31 00 43.09&16.93& 0.60& 60&0.43\\
PCG 1525+2956&  &           &        &               &            &   &   &  \\
A&15 25  2.97&+29 55 51.78&16.17& 0.30&-06&0.22\\
B&15 25  2.04&+29 55 56.96&16.34& 0.38& 69&0.07\\
C&15 25  2.21&+29 56 21.41&16.63& 0.53& 25&0.23\\
D&15 25  1.41&+29 55 54.98&17.59& 0.68& 12&0.05\\
PCG 1525+2824&  &           &        &               &            &   &   &  \\
A&15 25 21.66&+28 24 50.54&16.56& 0.60& 42&0.12\\
B&15 25 18.79&+28 24 41.68&17.27& 0.53&-42&0.26\\
C&15 25 21.12&+28 24 50.58&17.66& 0.36&-40&0.21\\
D&15 25 22.13&+28 25  2.71&18.38& 0.58& 85&0.34\\
PCG 1528+4235&  &           &        &               &            &   &   &  \\
A&15 28 51.56&+42 35 39.86&16.72& 0.21&-16&0.16\\
B&15 28 52.39&+42 35 57.92&17.72& 0.33& 28&0.25\\
C&15 28 54.21&+42 35 40.52&17.90& 0.11& 45&0.12\\
D&15 28 54.58&+42 36  0.83&17.97& 0.28& 21&0.51\\
E&15 28 54.22&+42 35 28.24&18.43& 0.29&-60&0.54\\
 &           &            &     &     &   &    \\
 &           &            &     &     &   &    \\
 &           &            &     &     &   &    \\
 &           &            &     &     &   &    \\
PCG 1533+4938&  &           &        &               &            &   &   &  \\
A&15 33 24.80&+49 39  3.53&16.73& 0.63& 71&0.18\\
B&15 33 24.90&+49 38 56.48&17.18&-0.02&-75&0.32\\
C&15 33 23.96&+49 38 42.41&17.75& 0.41& 42&0.16\\
D&15 33 22.51&+49 38 44.45&18.26& 0.38& 00&0.36\\
E&15 33 21.27&+49 39  6.67&18.39& 0.15& 77&0.28\\
PCG 1537+4507&  &           &        &               &            &   &   &  \\
A&15 37  9.96&+45 07 49.01&16.37& 0.43& 64&0.31\\
B&15 37 12.40&+45 08  1.43&16.56& 0.15& 59&0.62\\
C&15 37 13.32&+45 08  9.39&17.04& 0.35&-38&0.10\\
D&15 37 11.16&+45 08 18.16&17.72& 0.36& 06&0.32\\
PCG 1544+4125&  &           &        &               &            &   &   &  \\
A&15 44 44.78&+41 25 14.12&16.94& 0.23&-65&0.61\\
B&15 44 43.78&+41 25 15.96&17.90& 0.27&-25&0.41\\
C&15 44 43.77&+41 25  4.32&18.43& 0.29&-79&0.20\\
D&15 44 42.61&+41 25  6.46&18.73& 0.46& 00&0.27\\
PCG 1549+2756&  &           &        &               &            &   &   &  \\
A&15 49 30.29&+27 56 20.47&16.42& 0.02& 48&0.06\\
B&15 49 29.11&+27 56 41.78&17.48& 0.45& 34&0.34\\
C&15 49 31.49&+27 56 41.00&17.52& 0.14&-84&0.48\\
D&15 49 31.41&+27 56 46.83&18.09& 0.02&-71&0.27\\
PCG 1552+3344&  &           &        &               &            &   &   &  \\
A&15 52  5.38&+33 44 25.34&16.54& 0.38&-41&0.21\\
B&15 52  5.91&+33 44 48.59&16.91& 0.47&-71&0.22\\
C&15 52  4.07&+33 44 44.95&17.85& 0.31& 50&0.54\\
D&15 52  2.65&+33 44 52.54&18.14& 0.48& 23&0.14\\
 &           &            &     &     &   &    \\
 &           &            &     &     &   &    \\
 &           &            &     &     &   &    \\
 &           &            &     &     &   &    \\
 &           &            &     &     &   &    \\
PCG 1555+4604&  &           &        &               &            &   &   &  \\
A&15 55 19.09&+46 04 30.36&16.60& 0.28& 68&0.55\\
B&15 55 22.21&+46 04  0.08&16.86& 0.40&-69&0.06\\
C&15 55 20.51&+46 04 45.55&17.28& 0.13&-89&0.29\\
D&15 55 23.70&+46 04 45.15&17.76& 0.35&-77&0.10\\
E&15 55 22.35&+46 04  9.11&18.12& 0.25& 50&0.12\\
PCG 1603+5221&  &           &        &               &            &   &   &  \\
A&16 03 41.39&+52 21  8.08&16.79& 0.29&-87&0.31\\
B&16 03 39.27&+52 21 17.06&17.06&-0.09&-64&0.26\\
C&16 03 41.69&+52 21 20.37&17.47& 0.15& 85&0.19\\
D&16 03 42.34&+52 20 56.97&18.27& 0.32&-89&0.15\\
PCG 1609+3111&  &           &        &               &            &   &   &  \\
A&16 09  5.17&+31 11 18.82&16.58& 0.36&-16&0.10\\
B&16 09  5.64&+31 10 54.59&16.78& 0.40& 27&0.02\\
C&16 09  6.84&+31 11  2.40&18.31& 0.23&-38&0.28\\
D&16 09  4.78&+31 11 13.45&18.43& 0.23& 53&0.38\\
E&16 09  4.16&+31 10 53.40&18.53&-0.09& 48&0.26\\
PCG 1617+2758&  &           &        &               &            &   &   &  \\
A&16 17 55.20&+27 58 31.77&16.69& 0.42&-24&0.07\\
B&16 17 54.09&+27 58 13.65&17.73& 0.44&-53&0.25\\
C&16 17 55.64&+27 58 41.19&17.98& 0.32&-42&0.15\\
D&16 17 54.11&+27 58 54.95&18.33& 0.59&-20&0.48\\
PCG 1618+5531&  &           &        &               &            &   &   &  \\
A&16 18 41.08&+55 31  3.32&16.29& 0.44& 39&0.25\\
B&16 18 39.12&+55 31 12.65&16.77& 0.46&-33&0.23\\
C&16 18 39.30&+55 31  2.64&17.12& 0.34&-22&0.26\\
D&16 18 43.67&+55 31  7.39&17.32& 0.13& 22&0.28\\
 &           &            &     &     &   &    \\
 &           &            &     &     &   &    \\
 &           &            &     &     &   &    \\
 &           &            &     &     &   &    \\
PCG 1624+5446&  &           &        &               &            &   &   &  \\
A&16 24 13.66&+54 46 39.81&16.30& 0.54& 59&0.32\\
B&16 24 11.94&+54 46 29.35&17.61& 0.41& 21&0.28\\
C&16 24 14.03&+54 47  9.12&17.72& 0.47&-20&0.13\\
D&16 24 11.43&+54 46 42.37&18.03& 0.49& 36&0.13\\
PCG 1651+3129&  &           &        &               &            &   &   &  \\
A&16 51 25.83&+31 29 46.53&16.59& 0.29& 10&0.08\\
B&16 51 22.92&+31 29 41.43&17.15& 0.55& 79&0.07\\
C&16 51 23.29&+31 29 53.48&17.18& 0.19&-27&0.44\\
D&16 51 22.76&+31 29 19.21&18.30& 0.34& 70&0.10\\
PCG 1657+3529&  &           &        &               &            &   &   &  \\
A&16 57 26.03&+35 29 56.58&16.54& 0.08& 88&0.01\\
B&16 57 26.60&+35 29 21.12&17.06& 0.44& 25&0.04\\
C&16 57 25.99&+35 29 16.97&17.63& 0.42& 69&0.05\\
D&16 57 26.10&+35 29 32.66&17.92& 0.47&-61&0.10\\
PCG 1658+3705&  &           &        &               &            &   &   &  \\
A&16 58  6.28&+37 06  7.20&16.48& 0.58&-30&0.31\\
B&16 58  5.90&+37 06  9.24&17.30& 0.31&-21&0.13\\
C&16 58  7.69&+37 05 53.58&18.04& 0.49& 64&0.13\\
D&16 58  4.36&+37 05 44.61&18.14& 0.56&-69&0.19\\
PCG 1702+3335&  &           &        &               &            &   &   &  \\
A&17 02 58.24&+33 34 52.32&16.51& 0.39& 65&0.13\\
B&17 02 57.19&+33 35  0.26&16.88& 0.36&-65&0.15\\
C&17 02 59.49&+33 34 54.05&17.05& 0.31& 76&0.17\\
D&17 02 59.72&+33 35 26.95&17.53& 0.15& 18&0.18\\
PCG 1702+3331&  &           &        &               &            &   &   &  \\
A&17 03  0.14&+33 32 12.88&16.59& 0.22&-18&0.36\\
B&17 02 59.81&+33 31 35.05&16.73& 0.31&-30&0.30\\
C&17 02 59.15&+33 31 49.04&17.51& 0.09&-69&0.19\\
D&17 02 57.21&+33 31 54.19&18.52& 0.40&-89&0.31\\
 &           &            &     &     &   &    \\
PCG 1704+2818&  &           &        &               &            &   &   &  \\
A&17 04 57.36&+28 18 33.81&16.66& 0.50&-50&0.17\\
B&17 04 58.02&+28 18  9.29&17.11&-0.24& 56&0.21\\
C&17 04 59.77&+28 18 42.95&17.31& 0.27&-61&0.59\\
D&17 04 56.59&+28 18 50.58&17.63& 0.26& 58&0.15\\
PCG 1707+4203&  &           &        &               &            &   &   &  \\
A&17 07  0.51&+42 02 53.41&16.68& 0.54&-18&0.12\\
B&17 07  1.27&+42 03 13.94&17.05& 0.42& 25&0.06\\
C&17 07  0.55&+42 03  8.90&17.71& 0.48& 37&0.15\\
D&17 06 59.52&+42 03 13.32&18.31& 0.29& 09&0.45\\
PCG 1711+3946&  &           &        &               &            &   &   &  \\
A&17 11  2.26&+39 46 30.07&16.56&-0.04& 33&0.11\\
B&17 11  1.65&+39 46 43.24&17.24&-0.10&-79&0.37\\
C&17 11  3.82&+39 46 47.35&17.49&-0.11&-85&0.26\\
D&17 11  3.68&+39 46 54.70&18.08&-0.14&-72&0.04\\
PCG 1715+3545&  &           &        &               &            &   &   &  \\
A&17 15  1.68&+35 45 45.81&16.20& 0.29& 16&0.49\\
B&17 15  2.26&+35 45 46.87&16.49&-0.05&-13&0.30\\
C&17 15  1.19&+35 45 52.89&17.07& 0.15& 54&0.19\\
D&17 15  2.27&+35 45  4.59&17.35&-0.07& 24&0.49\\
PCG 1726+3848&  &           &        &               &            &   &   &  \\
A&17 26 31.34&+38 47 53.08&16.98& 0.74&-80&0.32\\
B&17 26 30.66&+38 48 22.71&17.30& 0.50&-69&0.23\\
C&17 26 30.57&+38 48  7.52&18.18& 0.44& 33&0.35\\
D&17 26 32.21&+38 48  5.01&18.91& 0.51& 29&0.36\\
\enddata
\end{deluxetable}


\begin{thebibliography}{}
\bibitem[]{} Allam, S., Tucker, D. 2000, {AN}, 321, 101 
\bibitem[]{} Arp, H. 1966, \apjs, 14, 1 
\bibitem[]{} Barton, E., Geller, M., Ramella, M., Marzke, R. O., da
Costa, L. 1996, \aj, 112, 871 
\bibitem[]{} Barton, E., de Carvalho, R.R., and Geller, M. 1998, \aj, 116, 1573
\bibitem[de Carvalho et al. (1997)]{deC1997} de Carvalho, R. R., Ribeiro,
A., Capelato, H., Zepf, S. E. 1997, \apjs, 110, 1 
\bibitem[]{} Carlberg R., Yee ,H., Morris, S., Lin, H., Hall, P.,
Patton, D., Sawicki, M. and Shepherd, C., 2001, \apj, 563, 736
\bibitem[]{} Coziol, R., Ribeiro, A. L. B., Capelato, H. V., de
Carvalho, R. R. 1998a, \apj, 493, 563
\bibitem[]{} Coziol, R., de Carvalho, R. R., Capelato, H. V., Ribeiro,
A. L. B. 1998b, \apj, 506, 545
\bibitem[]{} Coziol, R., Iovino, A., de Carvalho, R. R. 2000, \aj, 120, 47
\bibitem[]{} Cowie, L., Hu, E. M. - and Songaila, H. 1995, \aj, 110, 1576 
\bibitem[]{} Diaferio, A., Geller, M., Ramella, M. 1994, \aj, 107, 868
\bibitem[]{} Focardi, P., and Kelm, B. 2002, A\&A, 391, 35 
\bibitem[]{} Gal, R.R., de Carvalho, R. R., Odewahn, S.C., Djorgovski,
S.G., Mahabal, A., Brunner, R.J., and Lopes, P.A.A. 2002, \aj,
accepted 
\bibitem[]{} Hernquist, L. Katz, N., Weinberg D. 1995, \apj, 442, 57 
\bibitem[]{} Hickson, P. 1982, \apj, 255, 382 
\bibitem[]{} Iovino, A. 2002, AJ, 124, 2471  
\bibitem[]{} Mamon, G. 1986,\apj, 307, 426
\bibitem[]{} Mamon, G. 1987, \apj, 321, 622
\bibitem[]{} Mendes de Oliveira, C., \& Hickson, P. 1991, \apj, 380, 30
\bibitem[]{} Nolthenius, R., \& White, S.D.M. 1987, \mnras, 235, 505
\bibitem[]{} Odewahn, S.C., Gal, R.R., de Carvalho, R. R.,Djorgovski,
S.G., Mahabal, A., Brunner, R.J., Stalder, B., and Lopes, P.A.A. 2002,
submitted to AJ
\bibitem[]{} Peebles, P.J.E. Principles of Physical Cosmology, Princeton Series
in Physics, 1993.
\bibitem[]{} Prandoni, I., Iovino, A., MacGillivray, H. T. 1994, \aj, 107, 1235 
\bibitem[]{} Press, W., Schechter, P. 1974, \apj, 187, 425 
\bibitem[Ribeiro et al98]{Rib98}Ribeiro, A., de Carvalho, R. R.,
Capelato, H., Zepf, S. E. 1998, \apj, 497, 72
\bibitem[]{} Rose, J. A. 1977, \apj, 211, 311
\bibitem[]{} Shakhbazian R. K. 1973, Astrofizika, 9, 495  
\bibitem[]{} Stephan, M. 1877, \mnras, 37, 334
\bibitem[]{} Vorontsov-Velyaminov B.A. 1959, Atlas and Catalog of
Interacting Galaxies, Vol. 1, Moscow: Sternberg Inst.
\bibitem[]{} Vorontsov-Velyaminov B.A. 1977, Atlas and Catalog of
Interacting Galaxies, Part II, A\&A Suppl, 28, 1
\bibitem[]{} West, M., Oemler, A., Dekel, A. 1989, \apj, 346, 539
\end{thebibliography}
\end{document}